\documentclass[prl,twocolumn,superscriptaddress]{revtex4-1}
\usepackage[dvipdfmx]{graphicx}
\usepackage{graphicx}
\usepackage{physics}
\usepackage{bm, amsmath, amssymb, braket}
\usepackage{times}
\usepackage{multirow}
\usepackage{ascmac}
\usepackage{amsthm}
\usepackage{float}
\usepackage{color}
\usepackage{comment}
\usepackage{mathtools}  % for math
\usepackage{mathrsfs} 
\usepackage[colorlinks=True,urlcolor=blue,linkcolor=blue,citecolor=blue]{hyperref}% add hypertext capabilities
\usepackage{xcolor}
\usepackage{multirow}
\usepackage{setspace}
\usepackage{subfigure}
\usepackage[toc,page]{appendix}

\begin{document}
\title{Spontaneously formed excitonic density wave with vortex-antivortex lattice in twisted semiconductor bilayers}

\author{Deguang Wu}
\affiliation{National Laboratory of Solid State Microstructures and Department of Physics, Nanjing University, Nanjing 210093, China}
\author{Yiran Xue}
\affiliation{National Laboratory of Solid State Microstructures and Department of Physics, Nanjing University, Nanjing 210093, China}
\affiliation{Department of Physics, University of Massachusetts, 710 North Pleasant Street, Amherst, Massachusetts 01003-9337, USA}

\author{Baigeng Wang}
\email{bgwang@nju.edu.cn}
\affiliation{National Laboratory of Solid State Microstructures and Department of Physics, Nanjing University, Nanjing 210093, China}
\affiliation{Collaborative Innovation Center of Advanced Microstructures, Nanjing University, Nanjing 210093, China}
\affiliation{Jiangsu Physical Science Research Center, Nanjing University, Nanjing 210093, China}

\author{Rui Wang}
\email{rwang89@nju.edu.cn}
\affiliation{National Laboratory of Solid State Microstructures and Department of Physics, Nanjing University, Nanjing 210093, China}
\affiliation{Collaborative Innovation Center of Advanced Microstructures, Nanjing University, Nanjing 210093, China}
\affiliation{Jiangsu Physical Science Research Center, Nanjing University, Nanjing 210093, China}
\affiliation{Hefei National Laboratory, Hefei 230088, People's Republic of China }

\author{D. Y. Xing}
\affiliation{National Laboratory of Solid State Microstructures and Department of Physics, Nanjing University, Nanjing 210093, China}
\affiliation{Collaborative Innovation Center of Advanced Microstructures, Nanjing University, Nanjing 210093, China}
\affiliation{Jiangsu Physical Science Research Center, Nanjing University, Nanjing 210093, China}

\begin{abstract}
Exciton condensation, characterized by uniform phase coherence across macroscopic length scales, has enabled the discovery of a variety of excitonic states, greatly enriching our understanding of correlated many-body physics. More exotic quantum phenomena are anticipated when the phase factor develops spatial dependence. However, whether excitonic condensates with spatially modulated phase profiles can emerge spontaneously remains an open question. In this work, we uncover novel forms of excitonic density waves featuring nontrivial phase patterns in twisted semiconductor bilayers. Remarkably, we show that kinetic frustration inherent to these systems stabilizes excitonic condensates arranged into a vortex-antivortex lattice. This represents a class of correlated states previously unknown in two-dimensional semiconductors, wherein the phase degrees of freedom of exciton condensates play a defining role. Such states spontaneously break both time-reversal and inversion symmetries, leading to non-reciprocal exciton transport—an effect we term the excitonic diode effect. Furthermore, we compute and identify characteristic impurity-induced states in these unconventional condensates, providing distinct signatures for their experimental detection.
%Excitonic condensation, exhibiting a uniform phase coherence at macroscopic length scale,  has led to the discovery of diverse excitonic states, greatly enriching our understanding of correlated many-body physics. More exotic quantum phenomena may emerge when the phase factor acquires spatial dependence. However, it remains an open question whether excitonic condensates with spatially modulated phase factors can be spontaneously formed. In this work, we reveal novel types of excitonic density waves hosting nontrivial phase patterns in twisted semiconductor bilayers. Remarkably, we demonstrate that kinetic frustration inherent to these systems stabilizes an excitonic condensate forming a vortex-antivortex lattice.  This state spontaneously breaks both time-reversal and inversion symmetry, leading to non-reciprocal exciton transport behaviors, termed as the excitonic diode effect. Moreover, we dentify characteristic impurity-induced states in these unconventional condensates, offering clear signatures for their experimental detection. Our results highlight the essential role of the phase degrees of freedom in exciton condensates, pointing to a class of correlated states previously unrecognized in two-dimensional semiconductors.
\end{abstract}

\maketitle
\emph{\color{blue}{Introduction.--}}  The phase factor of many-body wave functions stands as one of the most fundamental quantities in condensed matter physics, underpinning nearly all macroscopic quantum phenomena \cite{eaCornell}, including superconductivity \cite{GinzburgLandau1950,Bardeen1957} and superfluidity \cite{Landau1941,Bogoliubov1947}. In superconductors, a uniform phase factor encodes the quantum coherence of Cooper pairs over macroscopic distances. Furthermore, when the phase factor becomes non-uniform and spatially dependent, more intriguing quantum behaviors emerge, such as dissipationless supercurrents \cite{London1935}, the Josephson effect \cite{Josephson1962}, and flux quantization associated with vortices \cite{Abrikosov1957,Blatter1994}. Among these, vortices--topological defects characterized by phase winding--have attracted considerable attention due to their potential to host Majorana fermions \cite{Read2000,Kitaev2001,liangfua,Ivanov2001}. Vortices can be generated not only in type-II superconductors under an applied magnetic field but also emerge spontaneously in certain chiral superconductors \cite{knigavko1998,Volovik1999}. Recent theoretical studies on twisted graphene multilayers \cite{Gaggioli2025} and hybrid superlattices intercalated with chiral molecules \cite{Mandal2025} have predicted a new class of chiral superconductors with mixed pairing symmetries. These systems exhibit a vortex-antivortex lattice (VAL) \cite{Gaggioli2025} and display novel transport properties such as the superconducting diode effect \cite{jhua,ando2020,daido2022,kjiang,jinxinhu}, revealing previously unrecognized consequences of the phase factor.

In parallel with superconductors, condensates of electron-hole pairs, i.e., excitons, can form similar coherent quantum states \cite{Keldysh,Jerome, Eisensteina, Eisensteinb, njzhang,ldua,zsuna}. Recent experimental advances in highly tunable two-dimensional (2D) materials have provided a new platform for realizing novel excitonic phases. A variety of examples have been reported, including excitonic density waves (EDW) \cite{Chen1991, Bi2021, Zeng2023, Dong2025, Kumar2025, jxhui}, excitonic Mott insulators \cite{bgao,zlian,rxiong}, excitonic superfluids \cite{Balatsky2004,Yasen,Taniguchi,Cutshall}, excitonic supersolids \cite{Matuszewski2012,Julku} Wigner crystal excitons \cite{Wigner1934,jyyou, zguo,rqi}, many-body renormalized excitons \cite{Efimkin}, and granulated order \cite{hliu}, among others. Although these discoveries have advanced the field of excitonic physics, they primarily focused on condensates with uniform phase factors. This raises an intriguing question: could excitonic states with spatially modulated phase factors--such as a vortex-antivortex lattice--form spontaneously, and might such phase modulation give rise to new quantum phenomena in excitonic systems?

%In this Letter, we uncover exotic excitonic condensates with nontrivial phase textures in twisted semiconductor bilayers. In particular, we identify a novel class of excitonic density wave (EDW) that spontaneously forms a lattice of vortices and antivortices, analogous to the VAL state recently found in superconductors \cite{Gaggioli2025,Mandal2025}. Remarkably, this excitonic VAL state spontaneously breaks both time-reversal and inversion symmetries, giving rise to nonreciprocal exciton transport—an effect we term the excitonic diode effect (EDE). Moreover, the excitons formed here are spin-polarized triplets. Thus, similar to the ``spin superconductors" proposed in Ref.[XX], the exciton transport results in a net spin current, but in a non-reciprocal fashion. 

In this Letter, we reveal the emergence of exotic excitonic condensates featuring nontrivial phase textures that spontaneously organize into a lattice of vortices and antivortices, akin to the VAL state recently discovered in superconducting systems \cite{Gaggioli2025,Mandal2025}. Remarkably, this excitonic VAL state spontaneously breaks both time-reversal and inversion symmetries, leading to nonreciprocal exciton transport—a phenomenon we term the excitonic diode effect (EDE). Furthermore, the excitons in this system are spin-polarized triplets. Consequently, analogous to the ``spin superconductor" proposal in Ref.\cite{qfsun,zqbao}, the exciton transport here gives rise to a net spin current, albeit in a nonreciprocal manner.

%Specifically, we rigorously derive that a twisted correlated semiconductor bilayer with a partially filled conduction band share the same low-energy physics with a frustrated quantum spin model on a triangular moiré lattice. Combining density-matrix renormalization group (DMRG) \cite{soslund,Garcia,Cirac} and a fermionization approach \cite{Sedrakyan2015,Sedrakyan2017,Sedrakyan2020,Wang2022,Wang2022B,Yang2022}, we identify three spin orders in correspondence to three different types of EDWs characterized by uniform, staggered, and VAL phase factors, respectively. Notably, the excitonic VAL state displays exotic topological properties, including the EDE and fractional charges localized at the vortex centers. To detect and distinguish between these EDWs, we further consider a quantum impurity coupled to these EDWs. Interestingly, the impurity states exhibit unique spatial dependence in their energetics, offering a clear fingerprint characterizing these EDWs. Our findings reveal unconventional excitonic condensates with nontrivial phase patterns, advancing fundamental knowledge beyond previously studied excitonic phases.
% Through electric gating and twist-angle control, all three EDWs lie within experimental reach.

Specifically, we establish that a twisted correlated semiconductor bilayer with a partially filled conduction band shares the same low-energy physics as a frustrated quantum spin model on a triangular moir\'{e} lattice. By combining density-matrix renormalization group (DMRG) calculations \cite{soslund,Garcia,Cirac,srwhitea,srwhiteb,Stoudenmire,Schollwock} with a fermionization approach \cite{Sedrakyan2015,Sedrakyan2017,Sedrakyan2020,Wang2022,Wang2022B,Yang2022}, we identify three distinct spin orders, each corresponding to a different type of EDW characterized by uniform, staggered, and VAL phase factors, respectively. Remarkably, the excitonic VAL state exhibits exotic topological properties, including the EDE and fractional charges localized at vortex cores. To detect and distinguish among these excitonic states, we further investigate a quantum impurity coupled to each EDW. Notably, the resulting impurity states display a characteristic spatial dependence in their energy spectra, providing a clear fingerprint for identifying the different EDW orders. Our findings uncover unconventional excitonic condensates with nontrivial phase textures, thereby advancing the fundamental understanding beyond previously studied excitonic systems.

\begin{figure}
\includegraphics[width=\linewidth]{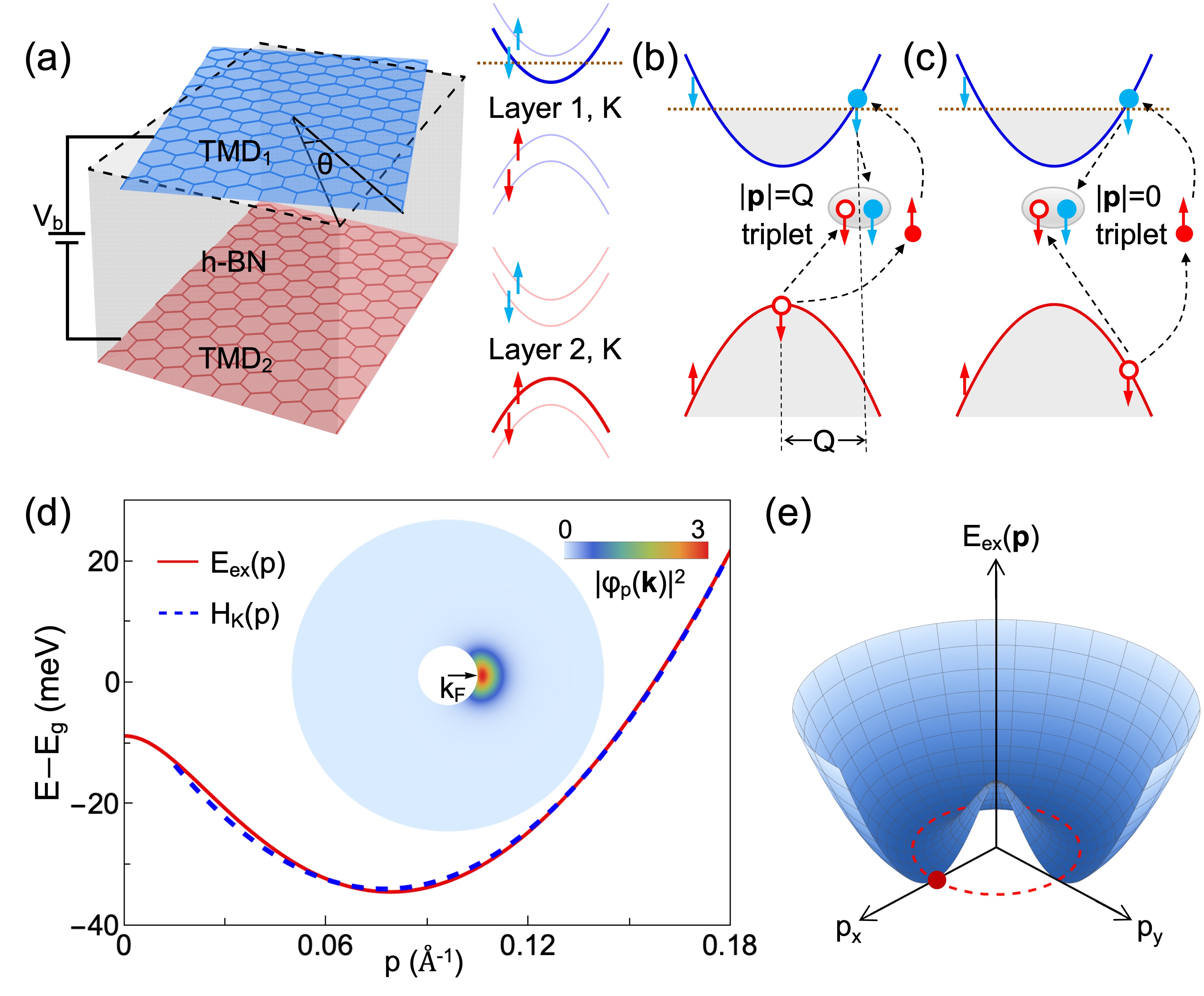}
\caption{\label{fig1}(a) A twisted TMD bilayer with electric gating. The band structure for the untwisted material is plot on the right for each layer. For type-II alignment, it  leads to a spin-porlarized direct gap  semiconductor for each valley, as shown in (b)(c). The low-energy excitons are formed by a spin-down hole in the valence band and a spin-down electron in the conduction band, leading to the spin-triplet pairing. When the conduction band is partially filled, the finite momentum excitons shown in (b) are more energetically favorable than the zero-momentum excitons depicted in (c).  (d) The calculated exciton dispersion displays a moat-like band as indicated by (e). The dashed curve in (d) is the fitting plot using Eq.\eqref{eq1}, and the inset to (d) shows the momentum space probability of the electron wavefunction, corresponding to the state marked by the red dot in (e). }
\end{figure}

\textit{\color{blue}{Interacting excitons on a moat band.--}}   We consider two transition metal dichalcogenides (TMD) monolayers separated by an insulating barrier, as shown in Fig.\ref{fig1}(a). The system can be controlled by both the electrostatic gating and the relative  twist angle between the two layers. For untwisted bilayers in the type-II alignment, the conduction band from the upper layer and the valence band from the lower layer form a spin-polarized direct gap semiconductor  \cite{Gupta2020,Jiang2021}, as shown in Fig.\ref{fig1}(b)(c). For each valley, the non-interacting low-energy states are described by, $H_0=\sum_{\alpha,\mathbf{k}}\varepsilon_{\alpha}(\mathbf{k})c^{\dagger}_{\alpha,\mathbf{k}}c_{\alpha,\mathbf{k}}$, where $c_{\alpha,\mathbf{k}}$ with $\alpha=c,v$ is the annihilation operator for the conduction and valence band electrons. Note that the spin index is omitted here, as it is locked to the band degrees of freedom.  The single-particle band is given by $\varepsilon_c(\mathbf{k})=\frac{\hbar^2k^2}{2m_c}-E_F$ and $\varepsilon_v(\mathbf{k})=-\frac{\hbar^2k^2}{2m_v}-E_g-E_F$, where $m_{\alpha}$ is the effective mass, $E_F$ is the Fermi energy, and $E_g$ is the band offset between the two layers that can be tuned by the displacement field. In the following, we focus on the case where the conduction band is partially filled with $E_F>0$.  We further consider the interlayer Coulomb interaction, which is responsible for exotic excitonic phases in TMD bilayers \cite{footnote1}. For the type-II alignment, it is cast in low-energy into the interband interaction, i.e., $H_{\mathrm{I}}=\frac{1}{2\mathcal{A}}\sum_{\mathbf{k},\mathbf{k}^{\prime},\mathbf{q}}V(\mathbf{q})c^{\dagger}_{c,\mathbf{k}+\mathbf{q}}c^{\dagger}_{v,\mathbf{k}^{\prime}-\mathbf{q}}c_{v,\mathbf{k}^{\prime}}c_{c,\mathbf{k}}$, where $V(\mathbf{q})=\frac{e^2}{2\epsilon_0\epsilon_r|\mathbf{q}|}e^{-d|\mathbf{q}|}$  with $\mathcal{A}$ denoting the system area.

%The model  $H=H_0+H_I$ describes a correlated semiconductor bilayer that has been extensively studied. In contrast, here we focus on two additional conditions, i.e.,  (1) the conduction (or valence band) is partially filled with a tunable $E_F$, and (2) there is a relative twist between the two monolayers. Then, we will explicitly show that various EDW states could be spontaneously  formed, including a novel vortex-antivortex lattice phase. 

In the case where the conduction band is partially filled, there occurs interesting features of the excitons. First, the excitons formed are spin-triplets, as indicated by Fig.\ref{fig1}(b)(c). Second, the particle-hole excitations with a finite total momentum, $|\mathbf{p}|=Q$, are most energetically favorable (Fig.\ref{fig1}(b)),  while the zero-momentum excitations have larger energy costs (Fig.\ref{fig1}(c)). This results in a moat-like dispersion for the excitons, as was initially reported in Ref.\cite{Cotlet2020}. Such a feature can be directly confirmed by variational calculations \cite{sup}. By minimizing the energy of excitons on top of the Fermi surface, i.e., $\langle \mathrm{FS}|\chi_{\mathbf{p}}H\chi^{\dagger}_{\mathbf{p}}|\mathrm{FS}\rangle$, where $\chi^{\dagger}_{\mathbf{p}}=\sum_{|\mathbf{k}|>k_F}\varphi_{\mathbf{p}}(\mathbf{k})c^{\dagger}_{c,\mathbf{k}}c_{v,\mathbf{k}-\mathbf{p}}$  is the exciton creation operator with  $\varphi_{\mathbf{p}}(\mathbf{k})$ being the electron wave function, the  exciton dispersion, $E_{\mathrm{ex}}(\mathbf{p})$, is obtained and shown in Figs.\ref{fig1}(d) and (e). It clearly exhibits a moat-like band with the energy minimum at $|\mathbf{p}|=Q$, and the kinetic energy is well described by \cite{Wang2023,Wang2024,liceran},
\begin{equation}\label{eq1}
    H_K(\mathbf{p})=\frac{\hbar^2(|\mathbf{p}|-Q)^2}{2m_{\mathrm{eff}}}+\Omega,
\end{equation}
where $m_{\mathrm{eff}}$ is the exciton effective mass and $\Omega$ is the energy minima measured from $E_g$. Correspondingly, the distribution of $|\phi_{\mathbf{p}=\mathbf{Q}}(\mathbf{k})|$ is plot in the inset of Fig.\ref{fig1}(d), which explicitly demonstrate the finite-momentum pairing of excitons.

In addition to the kinetic energy $H_K$, the dipolar interaction between excitons is non-negligible in TMD heterostructures. Taking into account the nonlocal screening effect \cite{Keldysh1979}, it is described by the Rytova-Keldysh potential \cite{Rytova1967}, 
\begin{equation}\label{eq2}
    V_{\mathrm{RK}}(|\mathbf{r}|)=\frac{\pi e^2}{\epsilon r_0}[H_0(\frac{|\mathbf{r}|}{r_0})-Y_0(\frac{|\mathbf{r}|}{r_0})],
\end{equation}
where $H_0(r)$ and $Y_0(r)$ are the Struve and Bessel functions of the second kind, respectively, $\epsilon$ is the dielectric constant, and $r_0$ is the screening length. Thus, the low-energy physics of the untwisted system is governed by $H_{\mathrm{ex}}=H_K(\mathbf{p})+\sum_{i<j}V_{\mathrm{RK}}(|\mathbf{r}_i-\mathbf{r}_j|)$, which describes interacting excitons on a moat band \cite{Wang2023}.

\begin{figure}
\includegraphics[width=\linewidth]{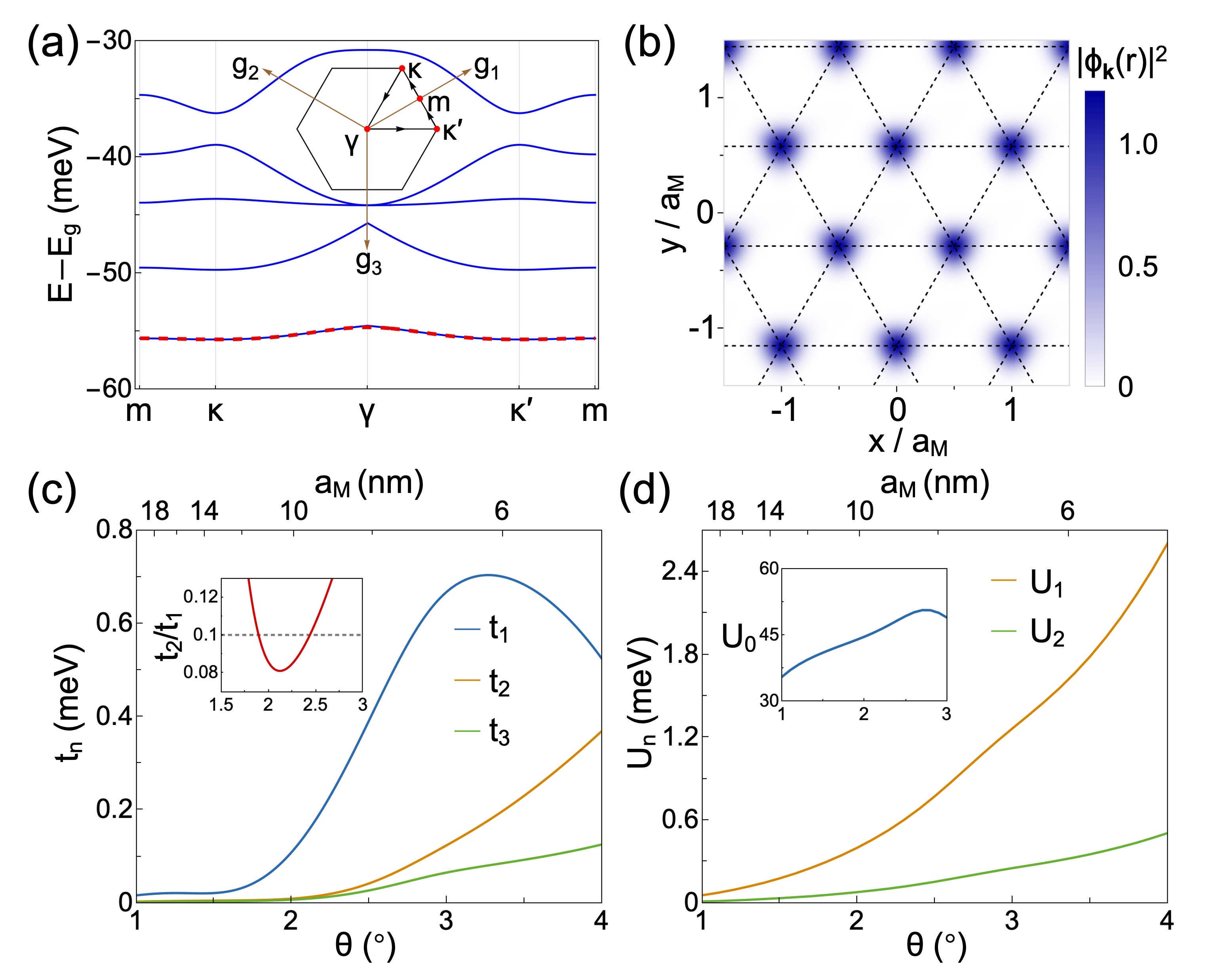}
\caption{\label{fig2}(a) This tight-binding model generates the single-particle dispersion as shown by the red dashed curve in Fig. 2(a), which well reproduces the dispersion of the lowest moire band. (b) The plot of the exciton wave function corresponding to the states on the lowest band in (a). (c) and (d) respectively show the calculated hopping amplitudes and interactions for the tight-binding model (Eq. \eqref{eq3}), as a function of the twist angle $\theta$. The inset to (c) shows the ratio $t_2/t_1$, which can be lower than 0.1 for $1.8^{\circ}\lesssim\theta\lesssim2.6^{\circ}$. The inset to (d) displays the on-site repulsion, which dominates over all other interactions. }
\end{figure}

\textit{\color{blue}{EDWs as analogs of spin orders.--}}  We now consider a relative twist between the two monolayers. Since the hole component is located at the valence band edge, the excitons are strongly modulated by a spatial moir\'{e} potential. In the first-order harmonic approximation, the moir\'{e} potential is of the form \cite{Wu2017, Wu2018, Wu2018B, Wu2019, Gotting2022, Devakul2021, Xie2024}, $\Delta(\mathbf{r})=2V\sum^3_{j=1}\cos{(\mathbf{g}_j\cdot\mathbf{r}+\phi)}$, where $\mathbf{g}_{j}$ denotes the moir\'{e} reciprocal lattice vectors satisfying  the threefold-rotational symmetry, and $V$ and $\phi$ characterize the amplitude and shape of the moir\'{e} potential, respectively. Taking into account both the kinetic and the moir\'{e} potential term, the moir\'{e} Bloch Hamiltonian of excitons can be expanded in the plane wave basis $|\mathbf{k}\rangle$ as, $\langle\mathbf{k}+\mathbf{g}^{\prime}|H|\mathbf{k}+\mathbf{g}\rangle=H_K(\mathbf{k}+\mathbf{g})\delta_{\mathbf{g}^{\prime},\mathbf{g}}+\Delta(\mathbf{g}^{\prime}-\mathbf{g})$, where
$\mathbf{k}$ is the wavevector in the mori\'{e} Brillouin zone. Diagonalization of $\langle\mathbf{k}+\mathbf{g}^{\prime}|H|\mathbf{k}+\mathbf{g}\rangle$ leads to the exciton bands shown in Fig.\ref{fig2}(a). Clearly, the moat-like dispersion in Fig.\ref{fig1}(d) is now flattened. Focusing on the lowest exciton band, the spatial distribution of the Bloch wave function $\tilde{\phi}_{\mathbf{k}}(\mathbf{r})$ is shown in Fig.\ref{fig2}(b), which implies that the excitons are confined at the moir\'{e} potential minima, forming a triangular moir\'{e} lattice. Correspondingly, the Wannier basis is constructed as, $w_{\mathbf{R}}(\mathbf{r})=\frac{1}{\sqrt{N}}\sum_{\mathbf{k}\in\mathrm{mBZ}}e^{-i\mathbf{k}\cdot(\mathbf{R}-\boldsymbol{\tau})}\tilde{\phi}_{\mathbf{k}}(\mathbf{r})$, where $\mathbf{R}$ denotes the site of the triangular lattice, and $\boldsymbol{\tau}$ specifies the position of a potential minimum in the moir\'{e} unit cell. 

The suppressed kinetic energy of the flattened exciton band highlights an important role of the exciton correlation $V_{\mathrm{RK}}$. Thus, we project $H_{\mathrm{ex}}$ onto the lowest moir\'{e} band in the Wannier basis \cite{sup}. This leads to an tight-binding model describing correlated excitons on a triangular lattice, i.e.,
\begin{equation}\label{eq3}
    H_{\mathrm{ex}}=\sum_{\mathbf{R},\mathbf{R}^{\prime}}[t(\mathbf{R}-\mathbf{R}^{\prime})b^{\dagger}_{\mathbf{R}}b_{\mathbf{R}^{\prime}}+\frac{1}{2}U(\mathbf{R}-\mathbf{R}^{\prime})b^{\dagger}_{\mathbf{R}}b^{\dagger}_{\mathbf{R}^{\prime}}b_{\mathbf{R}^{\prime}}b_{\mathbf{R}}],
\end{equation}
where $b^{\dagger}_{\mathbf{R}}$ ($b_{\mathbf{R}}$) is the creation (annihilation) operator for excitons, and $t(\mathbf{R}-\mathbf{R}^{\prime})$, $U(\mathbf{R}-\mathbf{R}^{\prime})$ are explicitly derived in Supplemental Materials. 

Denoting the $n$-th nearest neighbor hopping and interaction as $t_n$ and $U_n$ (with $U_0$ and $t_0$ being the onsite interaction and the chemical potential respectively), we plot $t_n$ and $U_n$ as a function of the twist angle $\theta$ in Figs.\ref{fig2}(c) and (d). As shown, for generic $\theta$, both $t_n$ and $U_n$ decay fast with $n$, with $t_1>t_2>t_3$, $U_0\gg U_1> U_2>0$, and $U_1$ being comparable with $t_1$. Moreover, $U_0$ dominants over all other terms, with $U_0\gg t_1$. Thus, the excitons are subjected to a very strong onsite repulsion that excludes the double occupancy states, making the excitons essentially hardcore bosons residing on the moir\'{e} lattice. In addition, in sharp contrast with the $E_F=0$ case that leads to $t_1<0$, the hopping amplitudes found here are all positive ($t_n>0$) for sufficiently large $E_F$. This is due to the finite-momentum pairing shown in Fig.\ref{fig1}(b),  essentially reflecting the kinetic frustration of the exciton moat band \cite{Wang2023}.

Keeping the dominant terms up to $n=2$, we obtain a reduced model, i.e.,
\begin{equation}\label{eq4}
\begin{split}
H^{\mathrm{\prime}}_{\mathrm{ex}}&=\sum_{n=1,2}\sum_{\mathbf{R},\nu}t_n(b^{\dagger}_{\mathbf{R}}b_{\mathbf{R}+\boldsymbol{\mu}^{n}_{\nu}}+h.c.)-\sum_{\mathbf{R}}\mu_bn_{\mathbf{R}}\\
&+\sum_{n=1,2}\sum_{\mathbf{R},\nu}\frac{U_n}{2}(n_{\mathbf{R}}-\frac{1}{2})(n_{\mathbf{R}+\boldsymbol{\mu}^n_{\nu}}-\frac{1}{2}),
\end{split}
\end{equation}
in addition with a hardcore constraint condition, $b^{\dagger2}_{\mathbf{R}}=b^2_{\mathbf{R}}=0$.  $n_{\mathbf{R}}=b^{\dagger}_{\mathbf{R}}b_{\mathbf{R}}$ is the exciton number operator, and $\boldsymbol{\mu}^1_{\nu}$ ($\boldsymbol{\mu}^2_{\nu}$) with $\nu=1,2,3$ are the three nearest (next nearest) neighbor moir\'{e} lattice vectors. This tight-binding model of hardcore excitons generates the dispersion as shown by the red dashed curve in Fig.\ref{fig2}(a), which well reproduces the dispersion of the lowest moir\'{e} band. The effective chemical potential $\mu_b$ here is a tuning knob relying on the interlayer bias voltage $V_b$ \cite{Ma2021, Gu2022}. By tuning $V_b$, signatures of excitonic insulators (EIs) have been experimentally reported \cite{Ma2021, Gu2022}, corresponding to the exciton condensates formed around $\mu_b\sim0$ in Eq. \eqref{eq4}. In the following, we will predict more exotic EDW states with delicate phase properties in this EI region.

Interestingly, representing hardcore bosons by spin operators, i.e., $S^-_{\mathbf{R}}=b_{\mathbf{R}}$, $S^+_{\mathbf{R}}=b^{\dagger}_{\mathbf{R}}$, and $S^z_{\mathbf{R}}=b^{\dagger}_{\mathbf{R}}b_{\mathbf{R}}-1/2$, Eq. \eqref{eq4} can be mapped to a spin-1/2 XXZ model on a triangular lattice for $\mu_b\sim0$, i.e.,
\begin{equation}\label{eq5}
H_{\mathrm{XXZ}}=\sum_{\mathbf{R},\nu,n=1,2}t_n(S^x_{\mathbf{R}}S^x_{\mathbf{R}+\boldsymbol{\mu}^{n}_{\nu}}+S^y_{\mathbf{R}}S^y_{\mathbf{R}+\boldsymbol{\mu}^{n}_{\nu}}+\lambda S^z_{\mathbf{R}}S^z_{\mathbf{R}+\boldsymbol{\mu}^{n}_{\nu}}),
\end{equation}
where we have let $\lambda=U_1/4_1\sim U_2/4t_2$ (Fig.\ref{fig2}(c)(d)) for simplicity. Since the parameters $t_n$ and $\lambda$ rely on specific experimental setups, including the twist angle, layer distance, dielectric screening, and the types of TMD materials, hereafter we treat them as tuning parameters for generality, and study all possible ground states of Eq. \eqref{eq5}.   

\begin{figure}
\includegraphics[width=\linewidth]{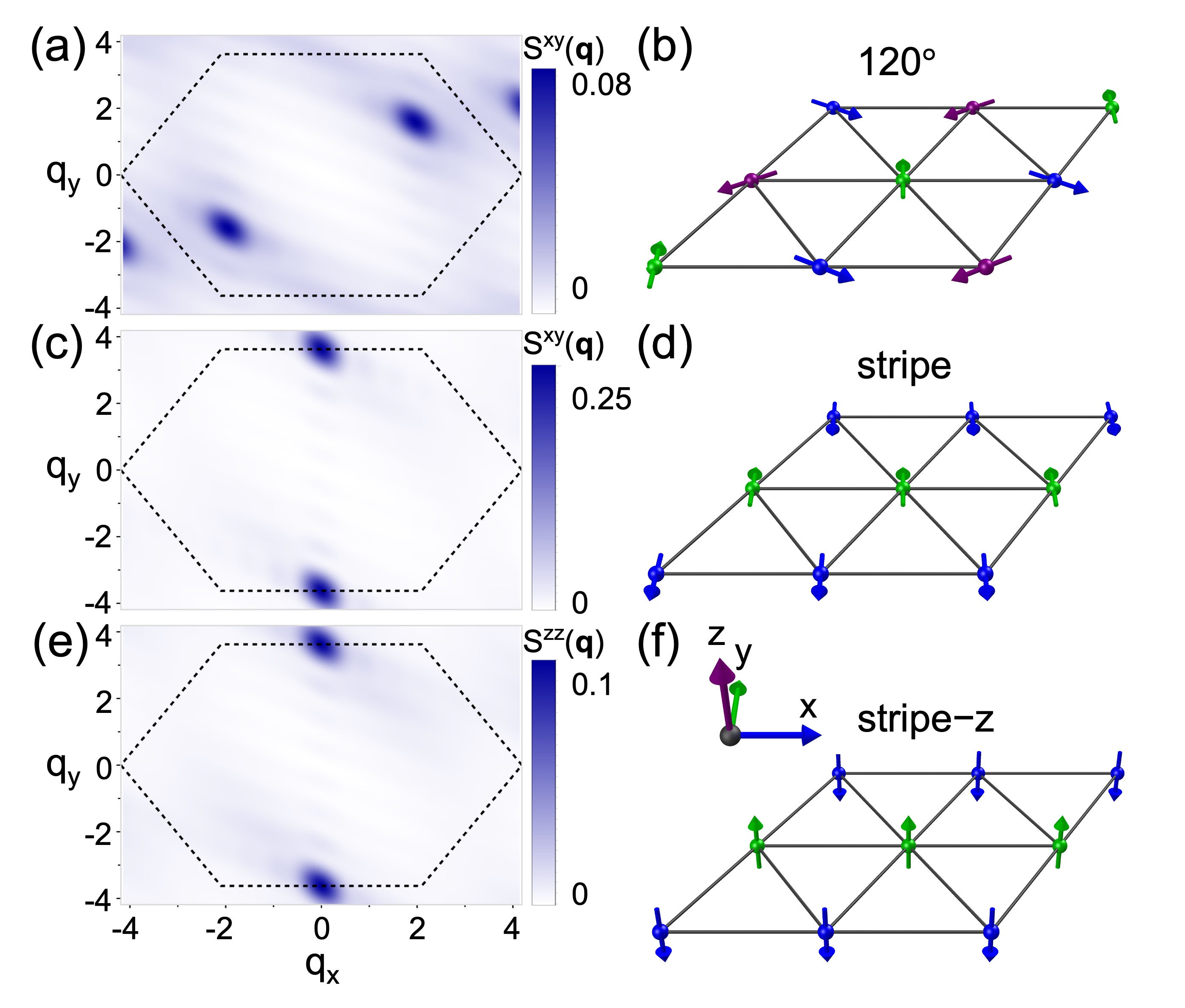}
\caption{\label{fig3}The DMRG results on the ground states corresponding to Eq. \eqref{eq5}, obtained on a finite sized cylinder of $L_x=10$ and $L_y=8$. (a)(c)(e) show the spin structure factors, i.e., $S^{\alpha\beta}(\mathbf{q})=\sum_{\mathbf{R},\mathbf{R}^{\prime}} \langle S^{\alpha}_{\mathbf{R}}S^{\beta}_{\mathbf{R}^{\prime}}\rangle e^{i\mathbf{q}\cdot(\mathbf{R}-\mathbf{R}^{\prime})}/N$ with $i,j=x,y,z$, for $(t_2/t_1, \lambda) = (0.06, 0)$, $(0.12, 0)$, and $(0.12, 1.2)$, respectively. The numerical details and parameters are included in Supplemental Materials. The peaks of $S^{ij}(\mathbf{q})$ reveal three distinct phases, i.e., the $120^{\circ}$ order, the in-plane stripe order and the stripe-z order, respectively. (b) illustrates the $120^{\circ}$ order, which prevails for small $\lambda$ and $t_2/t_1<0.1$. (d) depicts the in-plane stripe order, which stabilizes for small $\lambda$ and $t_2/t_1>0.1$. (f) displays the stripe-z order, which becomes dominant for $\lambda>1$. }
\end{figure}

We use the DMRG approach to simulate the ground state with varying $t_2/t_1$ and $\lambda/t_1$. For small $\lambda$ and $t_2$ with $\lambda<1$ and $t_2/t_1<0.1$, we arrive at a 120$^{\circ}$ spin order (Fig.\ref{fig3}(b)), as evidenced by the spin structure factor in Fig.\ref{fig3}(a). While for $t_2/t_1>0.1$ and $\lambda<1$, we obtain an in-plane stripe order (Fig.\ref{fig3}(c)(d)). In contrast, for $\lambda>1$ and  $t_2/t_1>0.1$, the stripe order shifts from in-plane to out-of-plane, forming a stripe-z order (Fig.\ref{fig3}(e)(f)). The above numerics can be further verified by mean-field calculations based on fermionization of the spin model \cite{sup}.  Using the method in Ref.\cite{Sedrakyan2015, Sedrakyan2017,Sedrakyan2020,Wang2022,Wang2022B,Yang2022}, we analytically confirm the above three phases as well as their phase transitions. Moreover, the fermionization approach clearly indicates a $p+ip$ pairing in the original exciton picture, as shown in Supplemental Materials.

\begin{figure}
\includegraphics[width=\linewidth]{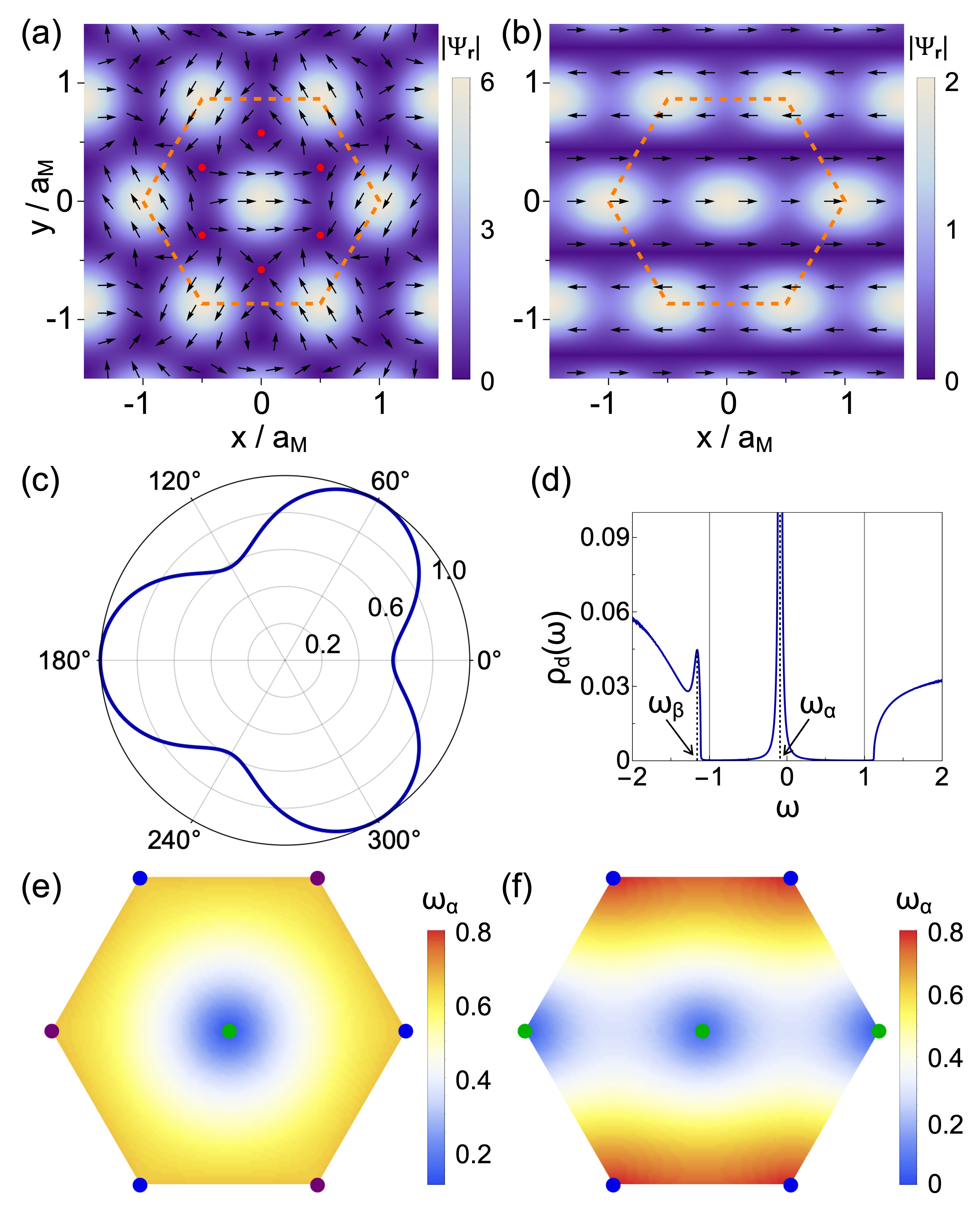}
\caption{\label{fig4} Demonstration of physical properties of the excitonic VAL state. (a) shows the real-space distribution of the condensate wave function $\Psi_{\mathbf{r}}$ for the VAL state. The color represents for its modulus while the arrows for its phase factor. The phase winding generates a VAL on a dual honeycomb lattice as denoted by the red dots. (b) shows the same as (a) but for the in-plane stripe EDW state. (c) The calculated exciton current density against the direction of bias field. (d) The impurity-induced YSR-like states for the excitonic VAL state, where the two generated impurity peaks are labeled by their energies, $\omega_{\alpha}$ and $\omega_{\beta}$. (e) and (f) are the color plot of $\omega_{\alpha}$ as a function of the impurity position for the excitonic VAL and the in-plane stripe EDW state, respectively. The plotted spatial region is marked by the dashed hexagon in (a) and (b).}
\end{figure}

\textit{\color{blue}{Excitonic VAL state.--}}  We now uncover the physical nature of the ``magnetic orders" in the original exciton picture. For the condensates, the excitonic local order parameter $\Psi_{\mathbf{r}}=|\Psi_{\mathbf{r}}|e^{-i\alpha_{\mathbf{r}}}$, satisfies $|\Psi_{\mathbf{r}}|^2\sim\langle b^{\dagger}_{\mathbf{r}}b_{\mathbf{r}}\rangle=\langle S^z_{\mathbf{r}}\rangle+1/2$. Hence, the stripe-z phase in Fig.\ref{fig3}(f) corresponds to the conventional EDW, exhibiting a periodic oscillation in the modulus $|\Psi_{\mathbf{r}}|$. Note that such an EDW has a coherent constant phase, i.e., $\alpha_{\mathbf{r}}=\alpha$. Furthermore,  the exciton wave function is related to the spin operators via, $\Psi_{\mathbf{r}}\sim\langle b_{\mathbf{r}}\rangle=\langle S^-_{\mathbf{r}}\rangle$ and  $\Psi^{\star}_{\mathbf{r}}=\langle b^{\dagger}_{\mathbf{r}}\rangle=\langle S^+_{\mathbf{r}}\rangle$.  Besides, $\langle S^-_{\mathbf{r}}\rangle=\langle S^x_{\mathbf{r}}\rangle-i\langle S^y_{\mathbf{r}}\rangle$  $=\langle S^p_{\mathbf{r}}\rangle e^{-i\beta_{\mathbf{r}}}$, where $\langle S^p_{\mathbf{r}}\rangle=\sqrt{\langle S^{x}_{\mathbf{r}}\rangle^2+\langle S^{y}_{\mathbf{r}}\rangle^2}$, and $\beta_{\mathbf{r}}=\mathrm{arctan}(\langle S^y_{\mathbf{r}}\rangle/ \langle S^x_{\mathbf{r}}\rangle)$ is the angle of the local spin expectation. Thus, it is clear that $\alpha_{\mathbf{r}}=\beta_{\mathbf{r}}$, i.e., the angle of the local spin order is identical to the phase of the excitonic condensates. From above, we know that the in-plane stripe phase displays a  $0$-$\pi$ staggered spatial modulation in $\alpha_{\mathbf{r}}$, as shown by Fig.\ref{fig4}(b).

Our main focus here is on the $120^\circ$ order. Since $\alpha_{\mathbf{r}}=\beta_{\mathbf{r}}$, the  $120^\circ$ order is translated into an EDW whose phase exhibits a periodic vortex-antivortex pattern, as explicitly plot by Fig.\ref{fig4}(a). The vortex cores form a honeycomb lattice,  dual to the original moir\'{e} triangular lattice, resulting in a VAL. Interestingly, the occurrence of the VAL has been recently found in the mixed-pairing superconductors in twisted graphene multilayers \cite{Gaggioli2025} and chiral molecule intercalated  hybrid superlattices \cite{Mandal2025}. However,  it has not been reported in any excitonic systems to date.  

The VAL here  is spontaneously generated due to the exciton frustration,  violating TRS and inversion symmetry (IS). As a result, such a phase  exhibits various novel quantum phenomena beyond conventional EIs. Similar to the superconducting counterpart  that hosts encircling inner supercurrent \cite{Abrikosov1957}, the VAL here drives circulating excitonic currents indicated by the arrows shown in Fig.\ref{fig4}(a). Furthermore, because the state breaks TRS and IS, non-reciprocal transport behaviors are expected. Introducing a thermodynamic field coupled to excitons \cite{footnote2}, we derive the response of the exciton current based on the Ginzberg-Landau theory \cite{sup} arising from Eq.\eqref{eq3}. As plot in Fig.\ref{fig4}(c), a strongly anisotropic exciton current is obtained, which sensitively relies on the field direction. The difference between the positive and negative current implies a significant non-reciprocal exciton transport. This is termed as the EDE, due to its analogy with the superconducting diode effect \cite{jhua,ando2020,daido2022,kjiang,jinxinhu}. Moreover, recall that the excitons are triplet and are spin-polarized \cite{qfsun,zqbao} as shown in Fig. \ref{fig1}(b), the  non-reciprocal exciton transport will be manifested as a non-reciprocal spin current, which is within experimental observation. We mention that the EDE found here is different in nature from the direction-dependent critical counterflow currents  recently predicted in spin-orbit-coupled electron-hole bilayers \cite{jxhui}, which originates from a finite-momentum condensates without displaying the VAL.

Another similarity with the superconducting VAL state \cite{Gaggioli2025} is that the excitonic pairing is of the $p+ip$ symmetry, as indicated by our fermionization mean-field theory \cite{sup}. Thus, the vortices would also host topological zero modes \cite{Wang2019}, akin to the Majorana fermions bound to the superconducting vortices \cite{Read2000,liangfua,Ivanov2001}. However, the key difference here is that the excitonic VAL  state resides in the electron basis, rather than the Nambu basis. Consequently, the zero modes here are no longer Majorana fermions, but are essentially solitons carrying fractionalized electron charges $e/2$ \cite{Wang2019,ywua,ywub}, which reside on a honeycomb lattice dual to the Mori\'{e} lattice.

Despite the above novel properties, it is intriguing to ask what is the  experimental signatures  of  the excitonic VAL state. Notably, recent  STM/STS experiments have revealed YSR-like \cite{Yu1965,Shiba1968,Rusinov1969} in-gap impurity states in excitonic insulators \cite{Yang2026,sKwon}, which can be utilized to distinguish different types of EDWs.  Thus, we consider a quantum impurity on top of EDWs at $\mathbf{r}_0$, which is coupled with the electrons and holes that are localized at the moir\'{e} lattice sites. As explicitly shown in Supplemental Material, we solve the effective impurity Green's function and obtain the local density of states at the impurity site, i.e., $\rho_d(\mathbf{r}_0,\omega)$, as a function of $\mathbf{r}_0$.

For generic $\mathbf{r}_0$, we generally arrive at YSR-like impurity states for the EDWs. As shown by Fig.\ref{fig4}(d), a pair of peaks emerge ($\alpha$ and $\beta$), and their energy positions ($\omega_{\alpha}$ and $\omega_{\beta}$) continuously evolve with the impurity location $\mathbf{r}_0$. In Fig.\ref{fig4}(e)(f), we plot $\omega_{\alpha}$ as a function of $\mathbf{r}_0$ for both the in-plane stripe and the excitonic VAL phase.  Interestingly, the two phases host completely distinct patterns regarding the impurity spectra. This originates from different interferences induced by the impurity hybridization, which are sensitive to the spatial distribution of the excitonic phases $\alpha_{\mathbf{r}}$. Therefore, the different patterns in Figs.\ref{fig4}(e) and (f) efficiently probe the key difference in the phase textures, offering an experimentally feasible signature of  the predicted EDWs.

\textit{\color{blue}{Discussion and conclusion.}}--We show that a novel EDW with VAL can be spontaneously formed in twisted semiconductor bilayers.  Since $\mu_b\sim0$ is most favorable for its observation, a proper electric control of the exciton chemical potential is desired. This can be experimentally achieved via electric gating in atomically thin TMD semiconductors \cite{Ma2021, Gu2022}, where the exciton binding energy is large and can be comparable with the quasi-particle gap.  Moreover, an appropriate twist angle $\theta$ is also indispensable. As shown by the inset to Fig.\ref{fig2}(c),  $1.8^{\circ}\lesssim\theta\lesssim2.6^\circ$ is expected to be the most promising parameter window. Notably, a tunable Bose–Hubbard system as described by Eq.\eqref{eq3} has been recently realized in doped moi\'{re} TMD bilayers \cite{rqi}, providing an ideal platform to further realize the excitonic VAL state.

Although the EIs and EDWs have been extensively studied for decades, their nontrivial phase effects are yet to be explored. The excitonic VAL revealed here contributes to a fundamentally new type of EDW exhibiting exotic phase effects. The emergent VAL breaks TRS and IS, generating non-reciprocal transport behaviors, e.g., the EDE, beyond all the previously studied excitonic systems. Moreover, a careful examination of the full phase diagram corresponding to Eq.\eqref{eq5} would further uncover an additional parameter region supporting a spin liquid phase, which is around $\lambda\sim1$ and $t_2/t_1\sim0.1$ \cite{Gallegos2025}. Translating the ``spin liquid" back into the excitonic picture will lead to an  excitonic topological order where the excitons avoid condensation but form highly entangled quantum liquid states, as originally proposed in Ref. \cite{Wang2023}.  Therefore, our work points to new possibilities to realize  correlated excitonic states beyond the  conventional EI framework, opening a new avenue towards a full understanding of bosonic quantum matter.

%Our fermionization mean-field theory suggests $p+ip$ wave pairing of the EDW. . Thus, such a topological EDW could support a spontaneous fractional charge lattice, which could be detectable by local STM/STS measurements. 

\begin{acknowledgments}
R. W. acknowledges Tigran Sedrakyan for fruitful discussions. This work was supported by the National Natural Science Foundation of China (No.12322402, No.12274206), the Scientific Research Innovation Capability Support Project for Young Faculty (SRICSPYF-ZY2025164), the National R\&D Program of China (2024YFA1410500, 2022YFA1403601), the Quantum Science and Technology-National Science and Technology Major Project (Grant No.2021ZD0302800), the Natural Science Foundation of Jiangsu Province (No.BK20233001), and the Fundamental Research Funds for the Central Universities (KG202501).

\end{acknowledgments}


\begin{thebibliography}{99}
\bibitem{footnote2} This can be realized by application of a temperature gradient or a laser field that generates a chemical potential bias of excitons, etc.
\end{thebibliography}
\end{document}

\begin{center}
\textbf{\large Supplemental material for}
\end{center}

\title{Spontaneously formed excitonic density wave with vortex-antivortex lattice in twisted semiconductor bilayers}

\author{Deguang Wu}
\affiliation{National Laboratory of Solid State Microstructures and Department of Physics, Nanjing University, Nanjing 210093, China}
\author{Yiran Xue}
\affiliation{National Laboratory of Solid State Microstructures and Department of Physics, Nanjing University, Nanjing 210093, China}
\affiliation{Department of Physics, University of Massachusetts, 710 North Pleasant Street, Amherst, Massachusetts 01003-9337, USA}

\author{Baigeng Wang}
\email{bgwang@nju.edu.cn}
\affiliation{National Laboratory of Solid State Microstructures and Department of Physics, Nanjing University, Nanjing 210093, China}
\affiliation{Collaborative Innovation Center of Advanced Microstructures, Nanjing University, Nanjing 210093, China}
\affiliation{Jiangsu Physical Science Research Center, Nanjing University, Nanjing 210093, China}

\author{Rui Wang}
\email{rwang89@nju.edu.cn}
\affiliation{National Laboratory of Solid State Microstructures and Department of Physics, Nanjing University, Nanjing 210093, China}
\affiliation{Collaborative Innovation Center of Advanced Microstructures, Nanjing University, Nanjing 210093, China}
\affiliation{Jiangsu Physical Science Research Center, Nanjing University, Nanjing 210093, China}
\affiliation{Hefei National Laboratory, Hefei 230088, People's Republic of China }

\author{D. Y. Xing}
\affiliation{National Laboratory of Solid State Microstructures and Department of Physics, Nanjing University, Nanjing 210093, China}
\affiliation{Collaborative Innovation Center of Advanced Microstructures, Nanjing University, Nanjing 210093, China}
\affiliation{Jiangsu Physical Science Research Center, Nanjing University, Nanjing 210093, China}

\def\tred#1{{\textcolor{red}{#1}}}
\captionsetup[figure]{name=Fig.}
\renewcommand{\thefigure}{S\arabic{figure}}
\renewcommand{\thetable}{S\arabic{table}}
\renewcommand{\theequation}{S\arabic{equation}}
% \captionsetup{justification=raggedright, singlelinecheck=false}

\begin{abstract}
We provide supplemental information regarding all technical details on the proofs and derivations of key conclusions presented in the manuscript.  Specific contents include: 1) Variational calculations leading to an excitonic moat band, 2) Derivation of the correlated Moir\'{e} exciton model using the Wannier basis, 3) Solving the correlated exciton model via the Chern-Simons fermionization approach, 4) Numerical details in DMRG calculations, 5) Detecting distinct excitonic density waves via impurity states, and 6) Excitonic diode effect.
\end{abstract}
\maketitle

\section{Variational calculations leading to an excitonic moat band}\label{Sec:1}

We employ a variational method to calculate the exciton dispersion in the absence of a moiré potential. The exciton creation operator is defined as
$\chi_{\mathbf{p}}^{\dagger}=\sum_{|\mathbf{k}|>k_\mathrm{F}}\varphi_{\mathbf{p}}(\mathbf{k})c_{c,\mathbf{k}}^{\dagger}c_{v,\mathbf{k}-\mathbf{p}}$, 
where $\mathbf{p}$ is the total exciton momentum and $\varphi_{\mathbf{p}}(\mathbf{k})$ represents the momentum space wave function of the constituent electron. The summation over the momentum $\mathbf{k}$ is restricted to $|\mathbf{k}| > k_\mathrm{F}$, which takes account into the Pauli blocking by the Fermi sea $|\mathrm{FS}\rangle$. To determine the kinetic energy of the exciton $E_{\mathrm{ex}}(\mathbf{p})$, we minimize the expectation value $W_{\mathrm{ex}}(\mathbf{p}) = \langle\mathrm{FS}|\chi_{\mathbf{p}}[H-E_{\mathrm{ex}}(\mathbf{p})]\chi_{\mathbf{p}}^{\dagger}|\mathrm{FS}\rangle$, where $E_{\mathrm{ex}}(\mathbf{p})$ acts as a Lagrange multiplier for normalization of the wave function. The variational condition $\delta W_{\mathrm{ex}}(\mathbf{p})/\delta \varphi_{\mathbf{p}}^*(\mathbf{k})=0$ leads to a Bethe-Salpeter-like equation,
\begin{equation}\label{seq1}
  [\varepsilon_c(\mathbf{k})-\varepsilon_v(\mathbf{k}-\mathbf{p})]\varphi_{\mathbf{p}}(\mathbf{k}) - \frac{1}{\mathcal{A}}\sum_{|\mathbf{k}^{\prime}|>k_\mathrm{F}}V(\mathbf{k}^{\prime}-\mathbf{k})\varphi_{\mathbf{p}}(\mathbf{k}^{\prime}) = E_{\mathrm{ex}}(\mathbf{p})\varphi_{\mathbf{p}}(\mathbf{k}),
\end{equation}
where $\varepsilon_{c(v)}$ is the conduction (valence) band energy and $V(\mathbf{q})$ is the interlayer Coulomb interaction.

For numerical convenience, we expand the electron wave function in an orbital angular momentum basis
\begin{equation}\label{seq2}
  \varphi_{\mathbf{p}}(\mathbf{k})  = \sum_{m \in \mathbb{Z}} \varphi_p(k,m)e^{im\theta},
\end{equation}
where $\theta = \theta_\mathbf{k} - \theta_\mathbf{p}$ is the angle of $\mathbf{k}=(k, \theta_\mathbf{k})$ relative to $\mathbf{p} = (p, \theta_\mathbf{p})$. Under this basis, Eq.~\eqref{seq1} is transformed into the radial form,
\begin{equation}\label{seq3}
  \left(\frac{k^2}{2m_{\mu}} + \frac{p^2}{2m_v} + E_g\right)\varphi_{p}(k,m) - 
  \frac{k\, p}{2m_v}[\varphi_{p}(k,m+1)+\varphi_{p}(k,m-1)]- \int_{k_\mathrm{F}}^{\Lambda} \frac{dk' k'}{2\pi}\tilde{V}(k,k',m)\varphi_{p}(k',m) = E_{\mathrm{ex}}(p)\varphi_{p}(k,m),
\end{equation}
where $m_\mu = m_c m_v / (m_c + m_v)$ is the reduced mass and $\Lambda$ is the momentum cutoff. Note that the $kp/m_v$ term arises from the expansion of the valence band energy $\varepsilon_v(\mathbf{k}-\mathbf{p})$, which couples to the adjacent angular momentum channels. In Eq. \eqref{seq3}, we used the fact that the interlayer Coulomb potential is diagonal in the angular momentum basis, i.e.,
\begin{equation}
  \frac{1}{(2\pi)^2} \iint V(|\mathbf{k}'-\mathbf{k}|) \, e^{im\theta - i m'\theta'} \, \mathrm{d}\theta \, \mathrm{d}\theta' 
  = \delta_{m,m'} \, \tilde{V}(k,k',m).
\end{equation}

We solve Eq.~\eqref{seq3} by discretizing the radial momentum $k$ on a uniform grid within $[k_\mathrm{F}, \Lambda]$ and truncating the angular momentum expansion at $|m| \le 5$. The exciton dispersion $E_{\mathrm{ex}}(p)$ is obtained from the lowest eigenvalue of the discretized matrix equation for each $p$ after numerical diagonalization. As shown in the main text, the wave function $|\varphi_{\mathbf{p}=\mathbf{Q}}(\mathbf{k})|$ is primarily concentrated around a finite momentum $\mathbf{Q}$. While we set $\theta_\mathbf{p} = 0$ in the inset of Fig. 1(c)  (aligning $\mathbf{Q}$ with the $k_x$-axis), Eq.~\eqref{seq2} demonstrates that the electron wavefunction distribution relies on  $\theta_\mathbf{p}$, and rotates simultaneously with $\mathbf{p}$. Due to the isotropic nature of the system, the energy $E_{\mathrm{ex}}(\mathbf{p})$ remains invariant under this rotation, depending only on the magnitude $p = |\mathbf{p}|$. This rotational symmetry leads to a degenerate ring of energy minima at finite momentum, resulting in the excitonic moat band.

\section{Derivation of the correlated Moir\'{e} exciton model using the Wannier basis}\label{Sec:2}

\begin{figure}[b]
   	\includegraphics[width=\linewidth]{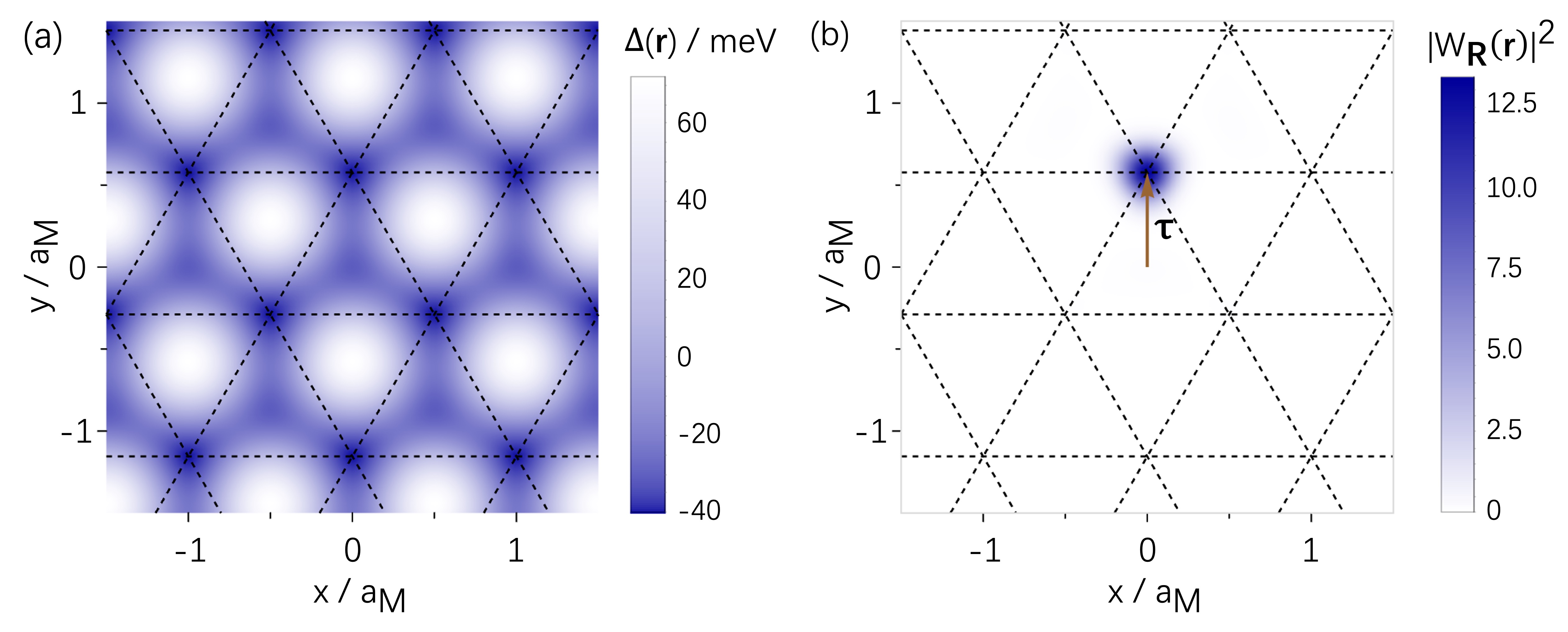}
   	\caption{(a) The exciton moir\'{e} potential landscape, where the potential minima form a triangular lattice. The parameters used are $V=12$ meV and $\phi=116^\circ$. (b) Spatial probability distribution of a localized exciton Wannier function centered at $\mathbf{R}=\boldsymbol{\tau}$.}\label{FigS1}
\end{figure}

In this section, we derive the effective correlated moir\'{e} exciton model (i.e., Eq.(3) of the main text) by constructing a Wannier basis for the lowest exciton miniband. First, the Hamiltonian consisting of the effective exciton kinetic energy $H_K$ and the moiré potential $\Delta(\mathbf{r})$ is numerically diagonalized using the plane-wave expansion method, with the summation over $\mathbf{g}$ truncated to ensure numerical convergence. For each momentum $\mathbf{k}$ in the moir\'{e} Brillouin zone (mBZ), the Bloch eigenstates are expanded as 
\begin{equation}\label{seq5} 
\phi_{n,\mathbf{k}}(\mathbf{r})=\sum_{\mathbf{g}} c_{\mathbf{g}}^{n,\mathbf{k}} e^{i(\mathbf{k}+\mathbf{g})\cdot\mathbf{r}}, 
\end{equation} 
where $c_{\mathbf{g}}^{n,\mathbf{k}}$ are determined by solving the matrix eigenvalue equation: 
\begin{equation}\label{seq6} 
\sum_{\mathbf{g}} \Bigl[H_K(\mathbf{k}+\mathbf{g})\delta_{\mathbf{g}',\mathbf{g}} + \Delta(\mathbf{g}'-\mathbf{g})\Bigr] c_{\mathbf{g}}^{n,\mathbf{k}} = E_n(\mathbf{k}) c_{\mathbf{g}}^{n,\mathbf{k}}. 
\end{equation} 

Focusing on the first moiré band ($n=1$), we construct maximally localized Wannier functions $W_{\mathbf{R}}(\mathbf{r})$. This requires smoothing the global phase of the Bloch states $\phi_{\mathbf{k}}(\mathbf{r})$ across the mBZ. As shown in Fig.~\ref{FigS1}(a), the moir\'{e} potential $\Delta(\mathbf{r})$ within a moir\'{e} unit cell reaches its minimum at the high-symmetry point $\boldsymbol{\tau} = a_M(0,1/\sqrt{3})$ (e.g., the MX stacking region). Since excitons are strongly localized at these minima, we choose a gauge where $\phi_{\mathbf{k}}(\boldsymbol{\tau})$ is real and positive for all $\mathbf{k}$. The gauge-fixed Bloch functions $\tilde{\phi}_{\mathbf{k}}(\mathbf{r}) = e^{i\psi_{\mathbf{k}}} \phi_{\mathbf{k}}(\mathbf{r})$, with $\psi_{\mathbf{k}} = -\mathrm{arg}[\phi_{\mathbf{k}}(\boldsymbol{\tau})]$, yield the Wannier function 
\begin{equation}\label{seq7} 
W_{\mathbf{R}}(\mathbf{r})= \frac{1}{\sqrt{N}}\sum_{\mathbf{k}\in\mathrm{mBZ}} e^{-i\mathbf{k}\cdot(\mathbf{R}-\boldsymbol{\tau})}\tilde{\phi}_{\mathbf{k}}(\mathbf{r}), 
\end{equation} 
where $W_{\mathbf{R}}(\mathbf{r})$ is centered at the potential minimum $\mathbf{R}=\boldsymbol{\tau}$ [Fig.~\ref{FigS1}(b)]. To fit the resulting moiré band structure, we employ a tight-binding Hamiltonian $H_{\mathrm{TB}}=\sum_{\mathbf{R},\mathbf{R}'}t(\mathbf{R}-\mathbf{R}')c_{\mathbf{R}}^{\dagger} c_{\mathbf{R}'}$, where the hopping amplitudes are extracted via 
\begin{equation}\label{seq8} 
t(\mathbf{R}-\mathbf{R}') = \frac{1}{N}\sum_{\mathbf{k}\in\mathrm{mBZ}}e^{i\mathbf{k}\cdot(\mathbf{R}-\mathbf{R}')}E(\mathbf{k}). 
\end{equation} 

To incorporate the strong many-body correlations characteristic of moiré systems, we now consider the exciton-exciton interaction. In the point-charge approximation, the bare dipole-dipole potential is $V_{\mathrm{dd}}(\mathbf{r}) = \frac{e^2}{4\pi\varepsilon_0\varepsilon_r} ( \frac{2}{r} - \frac{2}{\sqrt{r^2+d^2}} )$. For a more realistic treatment, we incorporate nonlocal dielectric screening via the Rytova–Keldysh potential $V_{\mathrm{RK}}(\mathbf{q}) = \frac{e^2}{2\varepsilon_0\varepsilon_r} \frac{1}{q(1+r_0q)}$. The screened dipole potential in momentum space thus becomes 
\begin{equation} 
    U(\mathbf{q}) = \frac{e^2}{\varepsilon_0\varepsilon_r} \frac{1-e^{-dq}}{q(1+r_0q)}. 
\end{equation} 

By projecting this interaction onto the Wannier basis, we obtain the interaction parameters $U_n \equiv U(\mathbf{R}-\mathbf{R}')$ as 
\begin{equation} 
U_{n}= \int \frac{d^{2}\mathbf{q}}{(2\pi)^{2}} U(\mathbf{q}) |M(\mathbf{q})|^{2} e^{i\mathbf{q}\cdot(\mathbf{R}-\mathbf{R}')}, 
\end{equation} 
where $M(\mathbf{q}) = \int d^2\mathbf{r} |W_{\mathbf{R}}(\mathbf{r})|^2 e^{-i\mathbf{q}\cdot\mathbf{r}}$ is the Wannier form factor, which captures the spatial extension of the localized exciton. The kinetic and interaction terms above altogether  define the correlated moiré exciton model, i.e., Eq.(3) in the main text.

\section{Solving the correlated exciton model via the Chern-Simons fermionization approach}\label{Sec:3}

We employ the Chern-Simons (CS) fermionization framework to provide a unified description of correlated exciton model, or equivalently, the spin-1/2 XXZ model on triangular lattice. Our analysis proceeds in two steps. First, in the easy-plane region, we consider $\lambda \ll 1$ where the Ising interaction is negligible, and focus on the competition between the $120^\circ$ order and the in-plane stripe order.
%We will show that, in the fermionized picture, the which arises from the interplay of low-energy Dirac fermion dispersion and gauge mediated interactions under distinct background flux configurations.
Then, we extend the theory to the easy-axis regime ($\lambda > 1$) by incorporating the Ising interaction. We will demonstrate how the density-density repulsions drive the system toward the out-of-plane magnetic order.

\subsection*{Chern-Simons Fermionization and Effective Low-Energy Theory}

Using the CS fermionization, we map the spin operators into spinless fermions by attaching a nonlocal string operator, i.e.,
\begin{equation}\label{seq11}
    S_{\mathbf{r}}^{\pm}=f_{\mathbf{r}}^{\pm} \exp \left[\pm i e\sum_{\mathbf{r}' \neq \mathbf{r}} \arg(\mathbf{r}-\mathbf{r}') n_{\mathbf{r}'}\right],
\end{equation}
where $f^+_{\mathbf{r}}=f_{\mathbf{r}}^{\dagger}$ and $f^-_{\mathbf{r}}=f_{\mathbf{r}}$  are respectively the creation and annihilation operator of spinless fermions, and $n_{\mathbf{r}} = f_{\mathbf{r}}^{\dagger} f_{\mathbf{r}}$ is the particle density operator, $e$ is the Chern-Simons charge which takes values of odd integers. An  advantage of this representation is that it respects the $SU(2)$ spin algebra without introducing hardcore condition (which is enforced by the Pauli exclusion of fermions).

\begin{figure}[t]
\includegraphics[width=\linewidth]{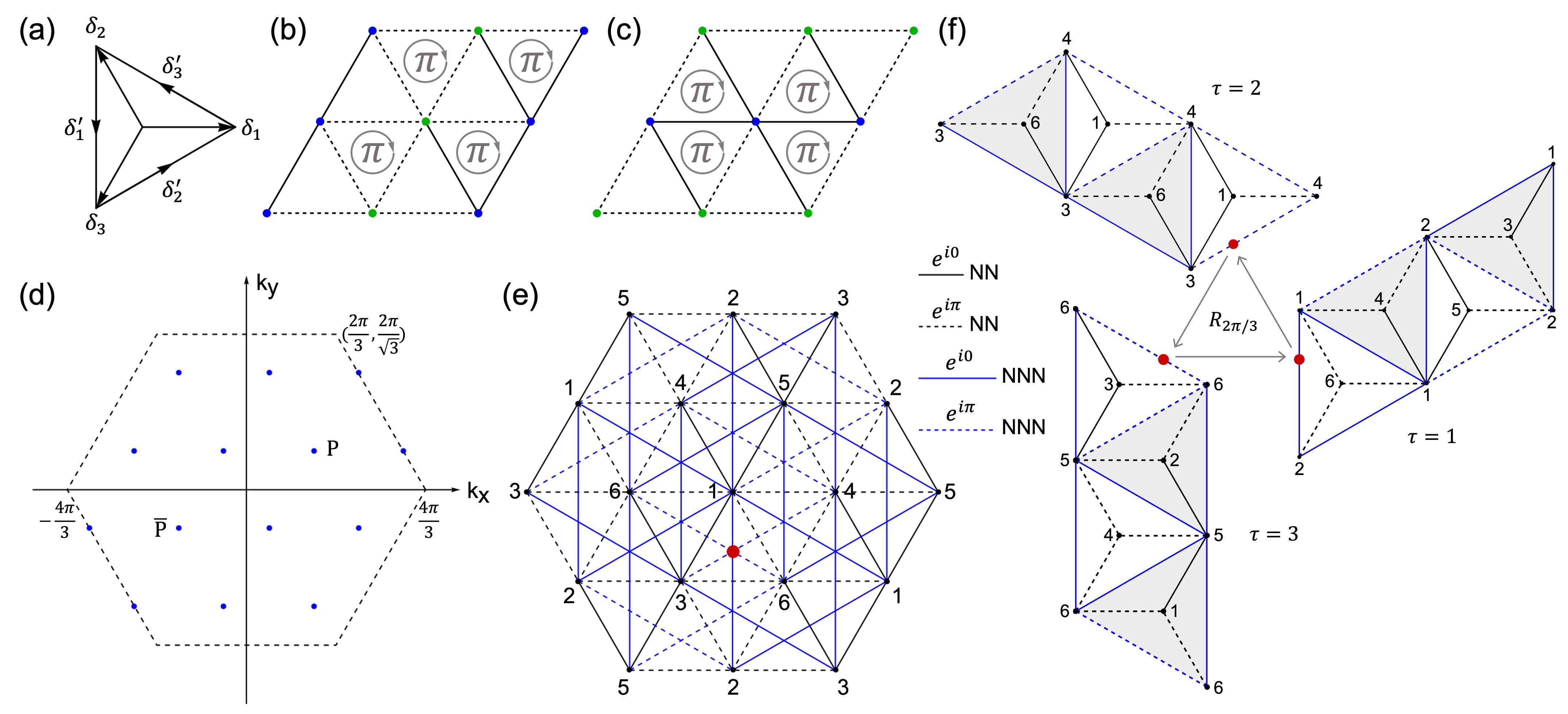}
\caption{(a) NN and NNN lattice vectors for triangular lattice. (b) and (c) are the plot of the $\pi$-flux patterns contributed by the NN gauge field for the 120$^\circ$ and stripe phase, respectively. The dashed bonds denote the hoppings with a sign flip compared to the solid bonds ($\pi$ phase shift), and the different site colors denote distinct sublattices. (d) shows the BZ of the original triangular lattice (the dashed lines). Marked in blue are the 12 Dirac points, corresponding to the low-energy fermionic theory describing the 120$^\circ$ phase.  $\mathbf{P}$ and $\mathbf{\bar{P}}$ are two representative Dirac points related by time-reversal operation. (e), (f) display the hopping configurations for the 120$^\circ$ phase. (e) depicts the hopping configuration that generates staggered $\pi$-fluxes, contributed by  both the NN (black) and  the NNN (blue) bonds. The numbers 1--6 label the six components of fermionic spinor. (f) shows the unit cells of the three NNN sublattices ($\tau = 1, 2, 3$) extracted from (e). The shaded triangles represent for the $\pi$-fluxes, and the red dots mark the $C_3$ rotation centers that relate the three configurations. }\label{FigS2}
\end{figure}

By substituting Eq.~\eqref{seq11} into the XXZ Hamiltonian (Eq.(5) of the main text) and neglecting the Ising-like interactions in the easy-plane region, we obtain an equivalent model describing spinless fermions coupled to a Chern-Simons $U(1)$ gauge field,
\begin{equation}
    H_{XY} = \sum_{n=1,2} \sum_{\mathbf{r}, \nu} t_n f_{\mathbf{r}}^{\dagger} f_{\mathbf{r}+\boldsymbol{\mu}_{\nu}^n} e^{i A_{\mathbf{r}, \mathbf{r}+\boldsymbol{\mu}_{\nu}^n}} + \text{H.c.},
\end{equation}
where $A_{\mathbf{r}_1, \mathbf{r}_2} = \sum_{\mathbf{r}} [\arg(\mathbf{r}_1-\mathbf{r}) - \arg(\mathbf{r}_2-\mathbf{r})] n_{\mathbf{r}}$ is the CS gauge field. Here, $\boldsymbol{\mu}_{\nu}^1=\boldsymbol{\delta}_{\nu}$ and $\boldsymbol{\mu}_{\nu}^2=\boldsymbol{\delta}'_{\nu}$, $\nu=1, 2, 3$, are the nearest-neighbor (NN) and next-nearest-neighbor (NNN) moiré lattice vectors for the triangular lattice, respectively [Fig.~\ref{FigS2}(a)]. This gauge field generates a magnetic flux of $B_{\mathbf{r}} = 2\pi n_{\mathbf{r}}$ threading per unit cell.

The CS gauge field is decomposed into a non-fluctuating background and fluctuations, $\mathbf{A}_{\mathbf{r}} = \overline{\mathbf{A}}_{\mathbf{r}} + \delta \mathbf{A}_{\mathbf{r}}$. In this subsection, we temporarily neglect the fluctuation $\delta \mathbf{A}_{\mathbf{r}}$, and derive the resulting low-energy effective theory under this approximation.  In the next subsection, we  will rigorously analyzed the effect of $\delta \mathbf{A}_{\mathbf{r}}$, which generates three ground states in consistence with the DMRG calculations presented in the main text.

At half-filling ($\langle n_{\mathbf{r}} \rangle = 1/2$), the mean-field background induces a modulated $\pi$-flux pattern. Specifically, one triangular plaquette of the unit cell is threaded by a $\pi$ flux, while the other remains flux free. Figures~\ref{FigS2}(b) and (c) illustrate NN $\pi$-flux configurations and the corresponding hopping sign structures for the $120^\circ$ and stripe phases, respectively. For the $120^\circ$ phase, interchanging the fluxes ($\pi \leftrightarrow 0$) leads to degenerate configurations, clearly implying a doubly degenerate ground states that are characterized by opposite helicities.

We now examine the $120^\circ$ phase with both NN and NNN hoppings. Its hopping structure is characterized by three disjoint sets of sublattices ($\tau=1,2,3$, and the sites of each sublattice are connected by NNN hoppings) exhibiting staggered $\pi$-flux, as illustrated in Figs.~\ref{FigS2}(e) and (f). To preserve the underlying $C_3$ symmetry, the flux distributions of these three sublattices are related by $R_{2\pi/3}$ rotations. The gauge field configurations shown above necessitate a six dimensional basis, $\Psi_{\mathbf{k}} = (f_{A,1}, f_{B,1}, f_{A,2}, f_{B,2}, f_{A,3}, f_{B,3})^T$, yielding a $6\times6$ Hamiltonian in the block matrix form as,
\begin{equation} 
H_0(\mathbf{k})= \begin{pmatrix}
\mathbf{h}_1(\mathbf{k}) & \mathbf{T}_{12}(\mathbf{k}) & \mathbf{T}_{13}(\mathbf{k}) \\
\mathbf{T}_{21}(\mathbf{k}) & \mathbf{h}_2(\mathbf{k}) & \mathbf{T}_{23}(\mathbf{k}) \\
\mathbf{T}_{31}(\mathbf{k}) & \mathbf{T}_{32}(\mathbf{k}) & \mathbf{h}_3(\mathbf{k})
\end{pmatrix}.
\end{equation}
The diagonal blocks, $\mathbf{h}_{\tau}(\mathbf{k}) = 2t_2 [ \cos(\mathbf{k} \cdot \boldsymbol{\delta}'_{1}) \sigma_x - \sin(\mathbf{k} \cdot \boldsymbol{\delta}'_{2}) \sigma_y + \cos(\mathbf{k} \cdot \boldsymbol{\delta}'_{3}) \sigma_z ]$, represent for the internal hopping within the $\tau$-th NNN sublattice. The off-diagonal blocks $T_{\tau, \tau'} = T_{\tau', \tau}^{\dagger}$, with $ \tau'=(\tau+1) \bmod 3$, denote the inter-flavor coupling mediated by the NN hopping $t_1$, i.e.,
\begin{equation}
    \mathbf{T}_{\tau, \tau'}(\mathbf{k}) = -t_1 (e^{-i\mathbf{k} \cdot \boldsymbol{\delta}_{1}} \sigma_x - i e^{-i\mathbf{k} \cdot \boldsymbol{\delta}_{2}} \sigma_y - e^{-i\mathbf{k} \cdot \boldsymbol{\delta}_{3}} \sigma_z).
\end{equation}

Despite this hybridization, the energy zeroes of the full system coincide with those of each individual lattice. The dispersion of the fermions is linear in low-energy, supporting multiple Dirac points as shown by Fig.~\ref{FigS2}(d). These Dirac points form a triangular lattice in momentum space. Taking a representative Dirac point $\mathbf{P}$, the effective low-energy theory around $\mathbf{P}$ is obtained by  expanding $H_0(\mathbf{k})$ as,  
\begin{equation}\label{seq15} 
    H_0^{(\mathbf{P})}(\mathbf{q}) = v_{\text{F}} \sum_{\nu} (\mathbf{q} \cdot \mathbf{c}_{\nu}) \sigma^{\nu},
\end{equation} 
where $\mathbf{c}_{\nu} = \boldsymbol{\delta}_{\nu} + t\ \boldsymbol{\delta}'_{\nu}$ ($t = t_2/t_1$), and $\sigma^{\nu}$ are the Pauli matrices acting in the sublattice space, and we have adopted $v_{\text{F}} = 2t_1 a_M$ as the energy unit. Around the time-reversed momentum $\mathbf{\bar{P}} = -\mathbf{P}$, the low-energy Hamiltonian can be obtained by $H_0^{(\mathbf{\bar{P}})}(\mathbf{q}) = [H_0^{(\mathbf{P})}(-\mathbf{q})]^*$, with $f_{\mathbf{q},\alpha} \to \bar{f}_{\mathbf{q},\alpha}$. Such an effective model gives rise to an isotropic Dirac spectrum $E_0(\mathbf{q}) = \pm v_0 |\mathbf{q}|$, with $v_0 = \sqrt{3(1 + 3t^2)/2}$. Note that Eq.\eqref{seq15} follows from the flux ansatz shown in Fig.~\ref{FigS2}(e), which we term the $120^{\circ}$ order flux ansatz. 

We then extend this CS fermionization approach to the stripe phase, which similarly features a $\pi$-flux configuration on both the NN and NNN sublattices. In the same six dimensional basis, the stripe phase Hamiltonian adopts a block matrix structure with blocks
\begin{align}
    \mathbf{h}_{\tau}(\mathbf{k}) &= 2t_2 \left[ \left( \cos(\mathbf{k} \cdot \boldsymbol{\delta}'_{2}) - \cos(\mathbf{k} \cdot \boldsymbol{\delta}'_{3}) \right) \sigma_x + \cos(\mathbf{k} \cdot \boldsymbol{\delta}'_{1}) \sigma_z \right], \\
    \mathbf{T}_{\tau, \tau'}(\mathbf{k}) &= t_1 \left[ \left( e^{i\mathbf{k} \cdot \boldsymbol{\delta}_{2}} - e^{i\mathbf{k} \cdot \boldsymbol{\delta}_{3}} \right) \sigma_x - e^{i\mathbf{k} \cdot \boldsymbol{\delta}_{1}} \sigma_z \right].
\end{align}
This leads to inversion symmetric Dirac points at $\pm\mathbf{Q}_{\tau} = [\mathrm{arccos}(t)+\frac{2\pi\tau}{3}, 0]$ with an anisotropic low-energy effective model
\begin{equation}\label{eqstrpeh}
    H_{0}^{(\mathbf{Q}_{\tau})}(\mathbf{q})=v_x q_x \sigma^3 + v_y q_y \sigma^1,
\end{equation}
where $v_x = \sqrt{1-t^2}$ and $v_y = \sqrt{3(1-t)/2}(1-2t)(1+t)$. Note that similar to the $120^{\circ}$ order discussed above, Eq. \eqref{eqstrpeh} follows from the  flux ansatz for the stipe phase, which demonstrates $\pi$-flux configurations contributed by both the NN and NNN bonds.

\subsection*{Gauge Field Induced Interaction and Phase Competition}

In this subsection, we incorporate the gauge field fluctuations $\delta \mathbf{A}_{\mathbf{r}}$ and obtain the mean-field ground states describing the $120^{\circ}$ order and the in-plane stripe order, as well as the phase transition between them. The Chern-Simons (CS) phases can be removed via a gauge transformation, which effectively replaces the momentum $\mathbf{q}=-i\boldsymbol{\nabla}$ by the covariant derivative $\mathbf{q}+e\mathbf{A}_{\mathbf{r}}$ (where $\mathbf{A}_{\mathbf{r}}$ hereafter denotes the fluctuation component for brevity). Or equivalently, one can integrate out the Chern-Simons gauge fluctuations. This will generate a nonlocal, effective fermion-fermion interaction. For the $120^\circ$ phase, this leads to an intervalley gauge field mediated interaction,
\begin{equation}\label{eq:Hint_full}
    H_{\mathrm{int}}=-\sum_{\mathbf{q},\mathbf{q}'}V_{\mathbf{q}-\mathbf{q}'}^{\alpha\alpha'\beta\beta'} f_{\mathbf{q},\alpha}^{\dagger}\bar{f}_{-\mathbf{q},\alpha'}^{\dagger}\bar{f}_{-\mathbf{q}',\beta}f_{\mathbf{q}',\beta'},
\end{equation}
where  $V_{\mathbf{q}-\mathbf{q}'}^{\alpha\alpha'\beta\beta'}=2\pi i e \left( \sigma_{\alpha\beta}^{\nu}\delta_{\alpha'\beta'}+\delta_{\alpha\beta}[\sigma^{\nu}]^{\text{T}}_{\alpha'\beta'} \right) \mathbf{A}_{\mathbf{q}-\mathbf{q}'}\cdot\mathbf{l}_{\nu}$, with $\mathbf{l}_{\nu}= \mathbf{e}_z\times\mathbf{c}_{\nu}=(3t\, \mathbf{a}_{\nu}-\mathbf{b}_{\nu})/\sqrt{3}$. Here, $\mathbf{A}_{\mathbf{k}}=\mathbf{k}/|\mathbf{k}|^2$ is the Fourier transform of the gauge potential. We then perform a Hubbard-Stratonovich transformation by introducing the CS pairing order parameters $\Delta_{\mathbf{q}}^{\alpha\alpha'}=-\sum_{\mathbf{q}'}V_{\mathbf{q}-\mathbf{q}'}^{\alpha\alpha'\beta\beta'} \langle\bar{f}_{-\mathbf{q}',\beta}f_{\mathbf{q}',\beta'}\rangle$. The interaction vertex functions are explicitly
\begin{gather}   
    V_{\mathbf{k}}^{1121}=V_{\mathbf{k}}^{1211}=V_{\mathbf{k}}^{1222}=V_{\mathbf{k}}^{2221}= 2\pi i e \mathbf{A}_{\mathbf{k}}\cdot\mathbf{l}^{-}, \notag \\
    V_{\mathbf{k}}^{1112}=V_{\mathbf{k}}^{2111}=V_{\mathbf{k}}^{2122}=V_{\mathbf{k}}^{2212}= 2\pi i e \mathbf{A}_{\mathbf{k}}\cdot\mathbf{l}^{+}, \notag \\
    V_{\mathbf{k}}^{1111}=-V_{\mathbf{k}}^{2222}=4\pi i e \mathbf{A}_{\mathbf{k}}\cdot\mathbf{l}_3,
\end{gather}
where $\mathbf{l}^{\pm}=\mathbf{l}_{1}\pm i\mathbf{l}_{2}$. 

 The order parameter can be further written in a more compact form $\Delta_{\mathbf{q}}^{\alpha\alpha'}=\Delta_{3\mathbf{q}}\sigma_{\alpha\alpha'}^0-i\Delta_{0\mathbf{q}}\sum_{\mu=1}^3\mathbf{q}\cdot\mathbf{l}_{\mu}\sigma_{\alpha\alpha'}^{\mu}$, which contains two order parameter components $\Delta_{0\mathbf{q}}$ and $\Delta_{3\mathbf{q}}$. It is important to note that the off-diagonal components $\Delta_{\mathbf{q}}^{12} \propto \mathbf{l}^{-}$ and $\Delta_{\mathbf{q}}^{21} \propto \mathbf{l}^{+}$ clearly exhibit the characteristic $p + i p$ symmetry, indicating a $p+ip$ order parameter. This pairing symmetry can be recast into the original exciton picture, which indicates a $p+ip$ pairing excitonic condensate.  

 For both the stripe and the $120^{\circ}$ order flux ansatz described above, the Bogoliubov-de Gennes (BdG) Hamiltonian, in the Nambu basis $\Psi_{\mathbf{q}}=[f_{\mathbf{q}A},f_{\mathbf{q}B},\bar{f}_{-\mathbf{q}A}^{\dagger},\bar{f}_{-\mathbf{q}B}^{\dagger}]^{\text{T}}$, is obtained in the following form,
\begin{equation}\label{eqbdg2}
    H_{\mathrm{BdG}}(\mathbf{q})=\sum_{\mu=1}^3 \left( \mathbf{q}\cdot\mathbf{c}_{\mu}\sigma^{\mu}\tau^3+\Delta_{0\mathbf{q}}\mathbf{q}\cdot\mathbf{l}_{\mu}\sigma^{\mu}\tau^2 \right) +\Delta_{3\mathbf{q}}\sigma^0\tau^1,
\end{equation}
yielding the quasi-particle spectra $\pm E_{\mathbf{q}}^a = \pm \sqrt{|(\Delta_{0\mathbf{q}}+a)v_0\mathbf{q}|^2+\Delta_{3\mathbf{q}}^2}$ for $a=\pm$. Note that the difference between the two flux ansatz only lie  in the different forms of $\mathbf{c}_{\mu}$ and $\mathbf{l}_{\mu}$ in Eq. \eqref{eqbdg2}.

By minimizing the ground state energy, $W=-\sum_{\mathbf{q},a}E_{\mathbf{q}}^a$, we obtain the self-consistent gap equations, i.e.,
\begin{equation}\label{eq:self_consistent_integrated}
    \Delta_{0\mathbf{q}}=\frac{e}{2} \sum_{a=\pm}\int_{0}^{q}\mathrm{d}q'\frac{q'\Delta_{3q'}}{q^2E_{q'}^a}, \quad \Delta_{3\mathbf{q}}=\frac{e v_0^2}{2} \sum_{a=\pm}\int_{q}^{\Lambda}\mathrm{d}q'\frac{(\Delta_{0q'}+a)q'}{E_{q'}^a}.
\end{equation}
For the Chern-Simons charge $e=1$, only the trivial solution exists, whereas our numerical solutions in Fig.~\ref{FigS3}(a) reveal that non-vanishing order parameters emerge for $e=3$, where the development of a finite gap stabilizes the magnetically ordered phase in the spin representation (more details about the Chern-Simons charge $e$ are included in Ref. [53-57] in the reference list of the main text). Solving the self-consistent gap equations for both flux ansatz (where Eq. \eqref{eq:self_consistent_integrated} corresponds to the $120^{\circ}$ order, while for the stripe phase, an identical form is obtained by momentum rescaling to remove anisotropy), we can further extract the ground state energy for the two phases, which is shown in Fig.~\ref{FigS3}(b). A transition between the two phases is identified, which is found to be located around $t_c=t_2/t_1 \approx 0.0956$. This is in well quantitive agreement with the DMRG results.

%Solving the self-consistent gap equations directly in the stripe phase is analytically challenging due to its strongly anisotropic Dirac dispersion. To circumvent this difficulty and establish a unified framework, we introduce a momentum rescaling: $\bar{\mathbf{q}} = v_0 \mathbf{q}$ for the $120^\circ$ phase and $(\bar{q}_x, \bar{q}_y) = (v_x q_x, v_y q_y)$ for the stripe phase. Following this transformation, both phases share an identical isotropic quasiparticle spectrum $E_{\bar{\mathbf{q}}}^a$ and identical scalar self-consistent equations. 

%The ground state energy is defined as $E_G = \frac{N_p}{|J|} \mathcal{W}$, where $N_p$ is the number of paired interaction vertices in the BZ that couple the original gapless modes, and $\mathcal{W} = \sum_{\bar{\mathbf{q}},a}-E_{\bar{\mathbf{q}}}^a$ is the universal rescaled energy sum. For the $120^\circ$ phase, the gauge fluctuations couple six pairs of time-reversal related modes ($N_p = 6$), whereas in the stripe phase, they couple four pairs of inversion related modes ($N_p = 4$). The Jacobian determinant $|J|$ maps the anisotropic cones to a circular form, with $|J| = v_0^2$ for the $120^\circ$ phase and $|J| = v_x v_y$ for the stripe phase. Since $\mathcal{W}$ depends exclusively on the summation of the rescaled quasiparticle spectrum, which is identical for both phases, the phase boundary is determined solely by the competition between the geometric factors $N_p$ and the Jacobians $|J|$, yielding the condition $2 v_0^2 = 3 v_x v_y$. 

\subsection*{Phase Transition from Stripe to Stripe-z Induced by Ising Term}

\begin{figure}[t]
\includegraphics[width=\linewidth]{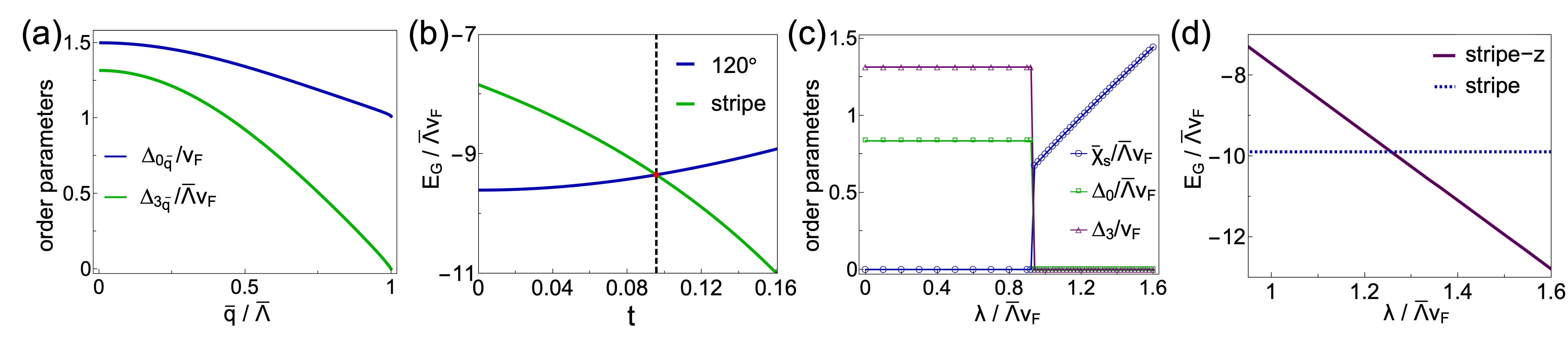}
\caption{(a) Numerical solutions for the CS superconducting order parameters, $\Delta_{0\bar{q}}$ and $\Delta_{3\bar{q}}$. The rescaled momenta are defined as $\bar{q} = v_0 q$ in the $120^\circ$ phase, and $(\bar{q}_x, \bar{q}_y) = (v_x q_x, v_y q_y)$ in the stripe phase. The CS charge is fixed at $e = 3$. (b) The calculated ground state energies of the $120^\circ$ and the stripe phases. The red dot marks the  transition point, $t_c \approx 0.0956$. (c) Evolution of the evaluated CS superconducting order parameters ($\Delta_0$, $\Delta_3$) and the excitonic insulating order parameter ($\bar{\chi}_s$), as a function of the Ising interaction strength $\lambda$. (d) The calculated ground state energies of the stripe and stripe-z phases across the phase transition driven by the Ising interaction.
}\label{FigS3}
\end{figure}

When $\lambda > 1$, the Ising interaction becomes dominant, driving a quantum phase transition from in-plane to out-of-plane magnetic order.  Considering the stripe-z phase predicted by DMRG, it is the NN Ising interaction that plays the essential role for its generalization.  In the CS fermionization representation, the dominant interaction here is given by $H_{\mathrm{Ising}}=\lambda\sum_{\mathbf{r}}n_{\mathbf{r},A}n_{\mathbf{r},B}$, which describes the density-density interaction between fermions on different sublattices, where $A$ and $B$ denote the two sublattices emerging in the stripe-z phase. Then, after projecting the interaction into the low-energy window around the Dirac points and performing a momentum rescaling, the interaction is reduced to,
\begin{equation}
    H_{\mathrm{Ising}} = \lambda \sum_{\bar{\mathbf{q}},\bar{\mathbf{q}}' , \bar{\mathbf{k}}} f_{\bar{\mathbf{q}} + \bar{\mathbf{k}}, A}^{\dagger} f_{\bar{\mathbf{q}}, A} f_{\bar{\mathbf{q}}' - \bar{\mathbf{k}}, B}^{\dagger} f_{\bar{\mathbf{q}}', B}.
\end{equation}
We then examine the low-energy physics in the band basis $(c_{\bar{\mathbf{q}},+}, c_{\bar{\mathbf{q}},-})^{\text{T}}$ via the  unitary transformation,
\begin{equation}
    \begin{pmatrix} f_{\bar{\mathbf{q}},A} \\ f_{\bar{\mathbf{q}},B} \end{pmatrix} = \begin{pmatrix} \cos\frac{\varphi}{2} & \sin\frac{\varphi}{2} \\ \sin\frac{\varphi}{2} & -\cos\frac{\varphi}{2} \end{pmatrix} \begin{pmatrix} c_{\bar{\mathbf{q}}, +} \\ c_{\bar{\mathbf{q}}, -} \end{pmatrix},
\end{equation}
where $\varphi$ is the polar angle of $\bar{\mathbf{q}}$. In this basis, the kinetic part $H_{\mathrm{str}}(\bar{\mathbf{q}})= \bar{q}_x \sigma_3 + \bar{q}_y \sigma_1$ is diagonal. Then, the transformation of $H_{\mathrm{Ising}}$ into the band basis yields two dominant forward scattering channels, $V^{(1)}$ and $V^{(2)}$, i.e.,
\begin{align}
    V^{(1)} &= -\lambda \sum_{\bar{\mathbf{q}}, \bar{\mathbf{q}}'} \cos^2\frac{\varphi}{2} \sin^2\frac{\varphi}{2}(c_{\bar{\mathbf{q}}, +}^{\dagger} c_{\bar{\mathbf{q}}, -} c_{\bar{\mathbf{q}}', -}^{\dagger} c_{\bar{\mathbf{q'}}, +} + \text{H.c.}), \notag \\
    V^{(2)} &= -\lambda \sum_{\bar{\mathbf{q}},\bar{\mathbf{q}}'} \left(\sin^4\frac{\varphi}{2} c_{\bar{\mathbf{q}}, +}^{\dagger} c_{\bar{\mathbf{q}}, -} c_{\bar{\mathbf{q}}', -}^{\dagger} c_{\bar{\mathbf{q}}', +} + \cos^4\frac{\varphi}{2} c_{\bar{\mathbf{q}}, -}^{\dagger} c_{\bar{\mathbf{q}}, +} c_{\bar{\mathbf{q}}', +}^{\dagger} c_{\bar{\mathbf{q}}', -}\right).
\end{align}
Within the mean-field approximation, we introduce the complex $s$-wave excitonic order parameter $\chi = |\chi|e^{\mathrm{i}\alpha} = \lambda \sum_{\bar{\mathbf{q}}}\langle c_{\bar{\mathbf{q}},+}^\dagger c_{\bar{\mathbf{q}},-} \rangle$, where $\alpha$ represents the U(1) phase. Restoring the pairing order parameters and transforming back to the sublattice $\otimes$ Nambu basis $\Psi_{\bar{\mathbf{q}}}$, the BdG Hamiltonian is then derived to be,
\begin{equation}\label{eq:BdG_Ising}
    \begin{aligned}
        H_{\mathrm{BdG}}(\bar{\mathbf{q}}) &= \left[ (\bar{q}_x - \bar{\chi}_s \cos\alpha \sin\varphi)\sigma_3 \right. \\
        &\quad \left. + (\bar{q}_y + \bar{\chi}_s \cos\alpha \cos\varphi)\sigma_1 + \bar{\chi}_s \sin\alpha \sigma_2 \right]\tau_3 \\
        &\quad + \Delta_{0\bar{q}}(\bar{q}_y\sigma_3 - \bar{q}_x\sigma_1)\tau_2 + \Delta_{3\bar{q}}\sigma_0\tau_1,
    \end{aligned}
\end{equation}
where $\bar{\chi}_s = \sum_a |\chi^{(a)}|$ represents for the sum of the amplitude of the order parameter at the Dirac valleys related by inversion symmetry. The resulting quasi-particle energy spectrum is then obtained as,
\begin{equation}
    E(\bar{\mathbf{q}}) = \pm \sqrt{\bar{q}^2(1+\Delta_{0\bar{q}}^2) + \bar{\chi}_s^2 + \Delta_{3\bar{q}}^2 \pm 2\Delta_{0\bar{q}}\bar{q}\sqrt{\bar{q}^2 + \bar{\chi}_s^2\sin^2\alpha}}.
\end{equation}
It is clear that the spectrum depends on the phase $\alpha$ if and only if $\Delta_{0\bar{q}} \neq 0$. Thus, this dependence may only occur around the transition point where  the in-plane orders and the out-of-plane order could coexist. In this case, there emerges a constraint on the phase $\alpha$, since  $\alpha = \pi/2$ is most energetically favorable in the mean-field ground state. Thereby, the self-consistent equation for the out-of-plane order is simplified to 
\begin{equation}
    \bar{\chi}_s = \lambda \sum_{a=\pm} \int_{0}^{\bar{\Lambda}} \mathrm{d}\bar{q} \frac{\bar{q} \bar{\chi}_s \left( 1 + a \frac{\Delta_{0\bar{q}}\bar{q}}{\sqrt{\bar{\chi}_s^2 + \bar{q}^2}} \right)}{E_{\bar{q}}^a}.
\end{equation}
Meanwhile, the self-consistent equations for $\Delta_{0\bar{q}}$ and $\Delta_{3\bar{q}}$ remain the same as derived above. We then numerically solve the coupled self-consistent equations for the three order parameters,  $\Delta_{0\bar{q}}$, $\Delta_{3\bar{q}}$, and $\bar{\chi}_s$. The solution indicate that for large $\lambda$, the out-of-plane order $\bar{\chi}_s$ dominates while the in-plane orders are suppressed to zero, as shown by Fig.~\ref{FigS3}(c). The corresponding mean-field ground state energy is also obtained and plot in Fig.~\ref{FigS3}(d),  signaling the transition towards the stripe-$z$ phase, in consistence with the DRMG simulations. 

\section{Numerical details in DMRG calculations}\label{Sec:4}

We study the $J_1$-$J_2$ model on the triangular lattice with Ising interaction using density matrix renormalization group (DMRG) approach based on matrix product states (MPS). The DMRG algorithm, originally introduced by \cite{Stevena,Stevenb} and has been successfully extended to quasi-two-dimensional systems \cite{Stoudenmire,Schollwock}. We map the triangular lattice onto a one-dimensional chain using a snake-like path with open boundary conditions (OBC) in both directions, which introduces effective long-range interactions in the MPS representation. To ensure sufficient accuracy, we
systematically choose the maximum bond dimension between $\chi_{\mathrm{max}} = 1000$ and up to $2000$, depending $J_2/J_1$ that tunes the  frustration of the system. Furthermore, we ensure that the
truncation error is lower than $10^{-5}$, such that the relevant entanglement of the ground state is accurately captured. 

In order to characterize different underlying phases, we evaluate the expectation $\langle S^{\alpha}_i\rangle$ and the spin-spin correlation function $\langle S^{\alpha}_iS^{\beta}_j\rangle$. From the correlation function, we compute the spin structure factor, which provides the direct information about the magnetic ordering. It is defined as
\begin{equation}
    S^{\alpha\beta}(\mathbf{q})=\frac{1}{N}\sum_{\mathbf{R},\mathbf{R}^{\prime}} \langle S^{\alpha}_{\mathbf{R}}S^{\beta}_{\mathbf{R}^{\prime}}\rangle e^{i\mathbf{q}\cdot(\mathbf{R}-\mathbf{R}^{\prime})},
\end{equation}
where $N$ is the total number of lattice sites, $S^{\alpha}_i$ denotes the $\alpha$-component of the spin operator at site $i$, and $\mathbf{r}_i$ represents for the position of site $i$. The results are shown in Fig.3 of the main text.

\section{Detecting excitonic density waves via impurity states}\label{Sec:5}

To elucidate the spectroscopic signatures of the EDW orders, we consider a local quantum impurity coupled to these excitonic condensate. We start from the real-space mean-field Hamiltonians describing the EDWs and consider the impurity problem on top of them. Taking the $120^\circ$ order, i.e., the VAL state as an example,  its real-space mean-field description is given by,
\begin{align}\label{eqsimph0}
H_0 = \sum_{\mathbf{k}} \sum_{s=1}^{3} \Big[ \varepsilon_{\mathbf{k}} c_{\mathbf{k},s}^{\dagger} c_{\mathbf{k},s} 
- \varepsilon_{\mathbf{k}} h_{\mathbf{k},s}^{\dagger} h_{\mathbf{k},s} 
+ \big( \Delta_s c_{\mathbf{k},s}^{\dagger} h_{\mathbf{k},s} + \text{H.c.} \big) \Big],
\end{align}
which contains an excitonic order parameter labeled by three emergent ``sublattices" denoted by $s$, in accordance with that of the $120^\circ$ order, such that the order parameter satisfies $\Delta_s = \Delta e^{i 2\pi (s-1)/3}$.  Correspondingly, the operators $c_{\mathbf{k},s}$ and $h_{\mathbf{k},s}$ annihilate electrons in the conduction and valence bands on sublattice $s$, respectively., and a  parabolic dispersions $\varepsilon_{\mathbf{k}} = \hbar^2 k^2 / 2m^*$ is taken for simplicity. The bare Green's function corresponding to Eq.\eqref{eqsimph0} (for the sublattice $s$)  is given in the Nambu basis $\Psi_{\mathbf{k},s} = (c_{\mathbf{k},s}, h_{\mathbf{k},s})^T$  by,
\begin{align}
\hat{G}^{(s)}_0(\mathbf{k},\omega) = \frac{1}{\omega^2 - E_{\mathbf{k}}^2}
\begin{pmatrix}
\omega + \varepsilon_{\mathbf{k}} & \Delta_s \\
\Delta_s^* & \omega - \varepsilon_{\mathbf{k}}
\end{pmatrix},
\end{align}
where $E_{\mathbf{k}} = \sqrt{\varepsilon_{\mathbf{k}}^2 + |\Delta|^2}$ is the quasi-particle dispersion.

Then, we consider  a quantum impurity localized at the position $\mathbf{r}_0$, which is described by $H_{\text{imp}} = \varepsilon_d d^{\dagger} d$, where  $\varepsilon_d$ is the impurity energy level.  The hybridization between the impurity and the EDW is naturally described by,
\begin{align}
H_{\text{hyb}} = \sum_{j} \big( v_j^{c} d^{\dagger} c_j + v_j^{h} d^{\dagger} h_j + \text{H.c.} \big),
\end{align}
which contains the hybridization with both the electron and hole components with strength $v^c_j$ and $v^h_j$, and the index $j$ here denotes the spatial coordinate $\mathbf{R}_j$. The hybridization strength decays  exponentially with the distance to the impurity site, thus takes the form as, $v_j^{\nu} \equiv v^{\nu}(\mathbf{R}_j - \mathbf{r}_0) = v_{\nu} e^{-|\mathbf{R}_j - \mathbf{r}_0|/\xi}$ for $\nu \in \{c, h\}$, where $\xi$ is the tunneling decay length. We introduce the ratio $\eta \equiv v_h / v_c$, which describes the hybridization asymmetry between the electron and the hole component. In momentum space, the hybridization is cast into,
\begin{align}
H_{\text{hyb}} = \frac{1}{\sqrt{N}} \sum_{\mathbf{k}, s} \big( \Lambda_{\mathbf{k},s}^c d^{\dagger} c_{\mathbf{k},s} + \Lambda_{\mathbf{k},s}^h d^{\dagger} h_{\mathbf{k},s} + \text{H.c.} \big),
\end{align}
with form factors $\Lambda_{\mathbf{k},s}^c = \sum_{\mathbf{R} \in s} v^c(\mathbf{R}-\mathbf{r}_0) e^{-i \mathbf{k}\cdot(\mathbf{R}-\mathbf{r}_0)}$ and $\Lambda_{\mathbf{k},s}^h = \eta \Lambda_{\mathbf{k},s}^c$.

The EDW state renormalizes the impurity Green's function by introducing a self-energy correction term, $\Sigma_d(\omega)$. Integrating out the EDW degrees of freedom yields
\begin{align}
\Sigma_d(\mathbf{r}_0,\omega) = \frac{1}{N} \sum_{\mathbf{k}, s} 
\begin{pmatrix} 
\Lambda_{\mathbf{k},s}^{c*} & \Lambda_{\mathbf{k},s}^{h*} 
\end{pmatrix} 
\hat{G}^{(s)}_0(\mathbf{k},\omega) 
\begin{pmatrix} 
\Lambda_{\mathbf{k},s}^{c} \\ 
\Lambda_{\mathbf{k},s}^{h} 
\end{pmatrix}.
\end{align}
Substituting the form factors and the Green's function components and utilizing $\Lambda_{\mathbf{k},s}^h = \eta \Lambda_{\mathbf{k},s}^c$, we obtain
\begin{align}\label{selfenergy}
\Sigma_d(\mathbf{r}_0,\omega) = \frac{1}{N} \sum_{\mathbf{k}, s} |\Lambda_{\mathbf{k},s}^c|^2 \frac{(\omega + \varepsilon_{\mathbf{k}}) + \eta^2 (\omega - \varepsilon_{\mathbf{k}}) + 2\eta \operatorname{Re}[\Delta_s]}{\omega^2 - \varepsilon_{\mathbf{k}}^2 - |\Delta|^2}.
\end{align}
Then, the $dI/dV$  spectra measured by STM on top of the impurity is proportional to the local density of states, i.e., $\rho_d(\mathbf{r}_0,\omega) = -\frac{1}{\pi} \operatorname{Im} [\omega - \varepsilon_d - \Sigma_d(\mathbf{r}_0,\omega + i0^+)]^{-1}$. 

From Eq. \eqref{selfenergy}, we know that the impurity state is sensitive to the phase factor pattern. This dependence comes in via the term $2\eta \operatorname{Re}[\Delta_s]$, as its values vary for different sublattices $s$. Thereby, as long as $\eta \neq 0$ (i.e., both the electron and hole component is hybridized to the impurity), the local $dI/dV$ spectra measured on top of the impurity will exhibit a ``sublattice" dependent feature. Moreover, the self-energy above also has dependence on the impurity position $\mathbf{r}_0$, which is implicit in the form factor $\Lambda^c_{\mathbf{k},s}$. Thus, Eq. \eqref{selfenergy} reflects the interference in the impurity hybridization, which detects the phase factor patterns of the EDWs. Note that the formalism above applies also for other types of EDW, not only the VAL state. Therefore, the impurity induced state would provide a clear experimental signature for their detection and distinction. 

%essential for distinguishing different EDW orders. The term couples the tip density of states directly to the local phase of the order parameter. 
%This formalism allows for the calculation of spatially resolved spectral maps, distinguishing between uniform, stripe ($ \Delta_A = -\Delta_B$), stripe-z ($\Delta_A = \Delta, \Delta_B = 0$), and $120^\circ$ EDW phases based on the periodicity of the spectral weight modulation.

\section{Excitonic diode effect}\label{Sec:6}

In this section, we derive the non-reciprocal transport properties, specifically the excitonic diode effect, originating from the VAL-EDW order on a $C_{3v}$ symmetric triangular lattice. We start from the tight-binding model of correlated bosons described in Eq.~(3) of the main text, and derive  a phenomenological Ginzburg-Landau (GL) free energy functional describing the low-energy physics of the VAL-EDW, from which the transport property can be obtained straightforwardly. 

From the $120^{\circ}$ order obtained in the equivalent spin picture (see main text), we know that the excitonic order parameter for this state is essentially  described  by a superposition of three modulation wave vectors $\mathbf{Q}_j$, namely,
\begin{equation}
\Psi(\mathbf{r}) = \Delta(\mathbf{r}) \sum_{j=1}^3 e^{i \mathbf{Q}_j \cdot \mathbf{r}}, \quad \text{with} \quad \mathbf{Q}_j = Q(\cos \tilde{\theta}_j, \sin \tilde{\theta}_j),
\end{equation}
where $\Delta(\mathbf{r})=1+\frac{1}{3}\sum_{i=1}^3\cos(\mathbf{g}_i\cdot\mathbf{r})$ describes the spatial localization of excitons on the Moiré lattice, $Q = |\mathbf{Q}_j|$, and $\tilde{\theta}_j = 2\pi(j-1)/3$ for $j=1,2,3$ in accordance with the $120^{\circ}$ order. This structure implies that the VAL ground state is characterized by simultaneous condensation of excitons at three distinct momenta in the BZ. The corresponding components in momentum space are denoted as $\psi_j$.

 In order to analyze the transport of excitons, we follow the standard method that extracts the supercurrent in a superfluid. Specifically, we consider the  momentum boost by $\mathbf{q}$, i.e.,  $\mathbf{k} \to \mathbf{Q}_j + \mathbf{q}$, and then examine the response in the free energy.  Particularly, it is the kinetic energy that is modified by such a momentum boost. Thus, the effective GL free energy density is obtained, taking the form as,
\begin{equation}
\mathcal{F} = \sum_{j=1}^3 \left[ \alpha_j(\mathbf{q}) |\psi_j|^2 + \beta |\psi_j|^4 \right],
\end{equation}
where $\beta > 0$ is a quartic coupling constant. The quadratic coefficient $\alpha_j(\mathbf{q})$ corresponding to the kinetic energy contribution is derived by expanding the tight-binding dispersion, $E(\mathbf{k}) = \sum_{\nu} [t_1\cos(\mathbf{k} \cdot \boldsymbol{\delta}_{\nu})+t_2\cos(\mathbf{k} \cdot \boldsymbol{\delta}'_{\nu})]$, around its minimum at the K point up to the third order. Notably, the expansion of $\alpha_j$ captures the $C_{3v}$ symmetry breaking via the trigonal warping term,
\begin{equation}
\label{eq:alpha_expansion}
\alpha_j(\mathbf{q}) = \alpha_0 + \alpha_2 |\mathbf{K}_j|^2 + \alpha_3 \left[ (\mathbf{K}_j)_x^3 - 3(\mathbf{K}_j)_x (\mathbf{K}_j)_y^2 \right],
\end{equation}
where $\mathbf{K}_j \equiv \mathbf{Q}_j + \mathbf{q}$. Here, $\alpha_0 < 0$ drives the ordering, $\alpha_2$ represents the effective mass term, and $\alpha_3$ characterizes the non-centrosymmetric trigonal warping.

Minimizing the free energy with respect to the order parameter amplitude, $\partial \mathcal{F} / \partial \psi_j^* = 0$, yields the equilibrium density, $|\psi_j|^2 = -\alpha_j(\mathbf{q})/(2\beta)$ for each component, provided $\alpha_j(\mathbf{q}) < 0$. Substituting this back into the free energy functional, we obtain the minimized free energy $\mathcal{F}_{\min}$ as a function of the transport momentum $\mathbf{q}$. The excitonic current $\mathbf{J}$, thermodynamically conjugate to the momentum shift, is then defined and calculated as
\begin{equation}\label{eqjq}
J(\mathbf{q}) \equiv \nabla_{\mathbf{q}} \mathcal{F}_{\min} = -\frac{1}{2\beta} \sum_{j=1}^3 \alpha_j(\mathbf{q}) \nabla_{\mathbf{q}} \alpha_j(\mathbf{q}).
\end{equation}
Following Eq.\eqref{eqjq}, one can also extract the current   along the direction of $\mathbf{q} = q(\cos\tilde{\theta}, \sin\tilde{\theta})$, i.e, $J(q, \tilde{\theta}) = \partial \mathcal{F}_{\min} / \partial q$.

By performing a small $q$ expansion and explicitly summing over $j$, the isotropic contribution dominates, while the interference between the $C_3$-symmetric order parameter and the trigonal warping generates a second-order anisotropic term. The current density is approximately obtained as,
\begin{equation}
\label{eq:J_approx}
J(q, \tilde{\theta}) \approx \mathcal{C}_1 q + \mathcal{C}_2 q^2 \cos(3\tilde{\theta}) + \mathcal{O}(q^3),
\end{equation}
where the coefficients are derived as,
\begin{align}
\mathcal{C}_1 &= -\frac{3}{4\beta} \left[ 4\alpha_0 \alpha_2 + 8\alpha_2^2 Q^2 + 16\alpha_2 \alpha_3 Q^3 + 9\alpha_3^2 Q^4 \right], \\
\mathcal{C}_2 &= -\frac{9\alpha_3}{4\beta} \left[ 2\alpha_0 + 8\alpha_2 Q^2 + 11\alpha_3 Q^3 \right].
\end{align}
The coefficient $\mathcal{C}_1$ here represents for the linear response (Drude-like weight), which is isotropic. The more important term is the second-order coefficient $\mathcal{C}_2$, which is non-zero if and only if $\alpha_3 \neq 0$.

The critical current is defined as the maximum current, i.e., $J_c(\tilde{\theta}) = \max_{q} |J(q, \tilde{\theta})|$. The presence of the $\cos(3\tilde{\theta})$ term in Eq.~(\ref{eq:J_approx}) breaks the reciprocity of the critical current. Specifically, under the inversion of current direction ($\tilde{\theta} \to \tilde{\theta} + \pi$), we have $\cos(3(\tilde{\theta} + \pi)) = -\cos(3\tilde{\theta})$. Consequently, $|J(q, \tilde{\theta})| \neq |J(q, \tilde{\theta} + \pi)|$, implying $J_c(\tilde{\theta}) \neq J_c(\tilde{\theta} + \pi)$. This confirms that the excitonic diode effect is a direct consequence of the VAL-EDW state, which breaks the effective inversion symmetry via the lattice trigonal warping.

%The origin of the diode effect lies in the coherent summation over the three $\mathbf{Q}_j$ vectors. While the lattice hosts the trigonal warping potential ($\alpha_3$), the macroscopic non-reciprocity manifests only because the $120^\circ$ order parameter locks into this lattice potential simultaneously along three directions. 